\documentclass[a4paper,12pt]{article}

\usepackage{amsmath,amssymb,graphicx}
\usepackage{pst-all}
\usepackage{cite}

\numberwithin{equation}{section}

\newcommand{\ii}{\mathrm{i}}

\newcommand{\dd}{\mathrm{d}}
\newcommand{\pd}{\partial}

\newcommand{\e}{\mathrm{e}}
\newcommand{\ket}[1]{\left|#1\right\rangle}
\newcommand{\bra}[1]{\left\langle #1\right|}
\newcommand{\bracket}[2]{\left\langle%
#1\left.\right|#2\right\rangle}

\newcommand{\tr}{\mathop{\mathrm{tr}}\nolimits}

\newcommand{\constant}{\mathrm{constant}}

\newcommand{\I}{\mathbb{I}}

\newcommand{\fc}{\check{\Phi}}
\newcommand{\ft}[2]{{\textstyle\frac{#1}{#2}}}
\newcommand{\qp}[1]{[\!\![ #1 ]\!\! ]}

\newcommand{\E}{\mathcal{E}}
\newcommand{\supp}{\mathop{\mathrm{supp}}}
\newcommand{\diag}{\mathop{\mathrm{diag}}}
\begin{document}

%\title{ Statistical mechanics for dilatations in
%$\mathcal{N}=4$
%super Yang--Mills theory}%
%\author{Corneliu Sochichiu\\
%{\it Max-Planck-Institut f\"ur Physik}\\
%{\it (Werner-Heisenberg-Institut)}\\
%{\it F\" ohringer Ring 6, D-80805 M\" unchen} \\
%e-mail: \texttt{sochichi@mppmu.mpg.de}
%}%

%
%\subjclass{}%
%\keywords{}%

%\date{}%
%\dedicatory{MPP-2006-95}%
%\commby{\hfill{MPP-2006-95}}%
% ----------------------------------------------------------------
\hfill{MPP-2006-95} %\maketitle
\\[1cm]
\begin{center}
  {\LARGE\bf Statistical mechanics for dilatations\\[0.3cm]
  in
$\mathcal{N}=4$ super Yang--Mills theory}\\[0.5cm]
{\large Corneliu Sochichiu\footnote{Post mail address after 01.01.2007: Laboratori Nazionali di Frascati, via Enrico Fermi 40, I-00044 Frascati (RM) Italy.}}\\[0.2cm]
{\it Max-Planck-Institut f\"ur Physik}
{\it (Werner-Heisenberg-Institut)}\\
{\it F\" ohringer Ring 6, D-80805 M\" unchen}
\\ and\\
 {\it Institutul de Fizic\u{a} Aplicat\u{a}}\\
 {\it str.Academiei, 5,
 MD-2028 Chi\c{s}in\u{a}u, MOLDOVA}
\\[0.5cm]
e-mail: \texttt{sochichi@mppmu.mpg.de}\\[0.5cm]
\end{center}

\begin{abstract}
 Matrix model describing the anomalous dimensions of composite operators
 in $\mathcal{N}=4$ super Yang--Mills  theory up to one-loop level
 is considered at finite
 temperature. We compute the thermal effective action for this
 model, which we define as the log of the partition function
 restricted to the states of given fixed length and spin. The result
 is obtained in the limit of high as well as low temperature.
\end{abstract}
% \maketitle
 \tableofcontents
% ----------------------------------------------------------------
\section{Introduction}

The gauge/string correspondence has a long history starting from the
't Hooft paper \cite{'tHooft:1973jz}, showing that the perturbative
expansion of a gauge theory can be organized according to the
geometrical genus expansion for the Feynman diagrams. The idea
behind this is that string theory may serve as an effective
description for a strongly interacting gauge system. The physical picture motivating such a description is given by condensation of the gauge field flux into tiny tubes leading to a linearly growing potential between quarks/antiquarks and string-like behavior.

The development of string theory lead to the conjecture of AdS/CFT correspondence \cite{Maldacena:1997re} (see the classical review on the subject \cite{Aharony:1999ti}). According to this conjecture the string theory can be described in terms of gauge fields which are nothing else that the collective coordinates of $D$-branes, the non-perturbative objects on which the fundamental strings can end. Therefore, the gauge field description of strings makes sense in the limit of strong string interactions.

In this picture the rank $N$ of the gauge group corresponds to the number of $D$-branes.
In the 't Hooft limit of large $N$ and small gauge coupling $g_{\rm YM}$ such that $g_{\rm YM}\sqrt{N}$ remains finite, the above duality
relation is reduced to the correspondence between non-interacting
IIB superstrings on AdS$_5\times S^5$ background and $\mathcal{N}=4$
supersymmetric Yang--Mills theory (SYM) on the Minkowski space which
is the conformal boundary to the anti-de Sitter space.

An important feature of the AdS/CFT correspondence is that it is a true i.e. Ising-type duality which means that the strong coupling dynamics is mapped to a weak coupling one and \emph{viceversa} (see \cite{RevModPhys.51.659} for a review of these type of dualities in lattice theories and spin systems). This property of AdS/CFT correspondence, beyond having a huge predictive force, also prevents from any direct proof of the correspondence itself, since for such a proof one should solve at least one of the models at strong coupling.

Last years a considerable progress was registered in the study of both string and gauge theories in the way of approaching the AdS/CFT conjecture. Among this an important point is the discovery of the integrability (see e.g.
\cite{Arutyunov:2004vx,Alday:2005gi,Minahan:2002ve,Beisert:2003jj}).

It was a breakthrough to realize that there are certain limits in which both string theory and SYM are reachable in the framework of perturbation theory. On the string theory side these limits are special geometrical limits while on the gauge theory side they correspond to particular subclasses of SYM composite operators for which one can extend the applicability of the perturbation theory
\cite{Berenstein:2002jq,Frolov:2002av,Frolov:2003xy}.

All this time the analysis was performed at infinite
$N$, which in particular allows one to use the integrability \cite{Minahan:2002ve,Beisert:2003jj,Beisert:2003tq}.
Since $1/N$ corrections are expected to break explicitly the
integrability, during this development they got much less attention.

In
\cite{Bellucci:2004ru,Bellucci:2004qx,Peeters:2004pt,Bellucci:2004dv,%
Bellucci:2005cu,Marrani:2006dp}, however, it was
shown that one can map the anomalous dimension operator on $\mathcal{N}=4$ SYM into an interacting spin system taking into account the finite $N$ rank of the gauge group. The finiteness of $N$, results in the chain splitting and joining interaction. This process mimics the string interaction.\footnote{To be precise, the above approach neglects the trace identities for U($N$) matrices. This means that only \emph{polynomial} $1/N$ contribution is taken into account and not the exponential corrections. This is very similar to the Quantum Field Theory, where non-analytic corrections are systematically dropped in perturbation theory.}

There is, however, still another approach to the same problem. It consists in the interpretation of the operator of anomalous dimensions as the Hamiltonian of a gauged matrix model \cite{Bellucci:2004fh,Agarwal:2004cb,Agarwal:2004sz}.  Spin chains in this picture correspond to gauge invariant collective states. The problem with such a description is that there are not enough developed tools for the study of such type of matrix models independently of the spin chain results.

One possibility is to describe the model in terms of a noncommutative gauge theory by expanding the matrix fluctuations about an appropriate classical solution \cite{Bellucci:2004fh}. This approach however, has its own drawbacks since it leaves a number of unanswered questions regarding the arbitrariness of such a description as well as regarding the spin chain/string interpretation. In particular, in the noncommutative description the matrix model of it is not very clear what is e.g. a spin chain.

The present paper aims to fill, at least partially, the gaps in the understanding of the matrix model for the anomalous dimensions of $\mathcal{N}=4$ SYM. Here we consider  statistically large but finite values of $N$. This means that we do not exclude the string interactions
from our description. According to BMN analysis \cite{Berenstein:2002jq}, the string interaction is rated as
$L^2/N$, where $L$ is the aggregate length of all spin chains. This interaction becomes important when one considers long enough operators.

In our approach we find it convenient to put the matrix theory in a finite temperature background. In the Yang--Mills theory this temperature corresponds to the Euclidean compactified Liouville or scale direction. According to AdS/CFT correspondence this is a compact time cycle in the AdS space. As it was pointed in \cite{Witten:1998zw}, the thermalization of AdS space occurs due to the presence of black holes.

The $\mathcal{N}=4$ SYM theory is a conformal invariant theory, therefore its dilatation operator can be identified with the Hamiltonian dual to the radial time. Since the dilatation operator and the SYM Hamilton are related through a similarity transformation, the spectrum and therefore, the thermodynamical properties of the matrix model should be the same as those of the thermal Yang--Mills which were considered in \cite{Sundborg:1999ue,Aharony:2003sx,Aharony:2005bq}.\footnote{The SU(2) sector can be regarded as a particular limit of $\mathcal{N}=4$ SYM \cite{Harmark:2006di,Harmark:2006ie,Harmark:2006ta}.} There is a difference, however. In the contrast to the above cited works where the thermal Yang--Mills was considered on $S^1\times S^3$, the radial time temperature should correspond rather to SYM compactified on $S^4$.

On the other hand, a similar partition function was considered in \cite{Polyakov:2001af,Bianchi:2003wx,Spradlin:2004pp}. In particular, in \cite{Spradlin:2004pp}, the thermal partition function of the gauged oscillator is derived by counting the trace states using the P\'olya Enumeration Theorem (PET). The problem of such approach is that, strictly speaking, it requires $N$ to be infinite. Because of this the analysis can not be extended beyond the Hagedorn temperature. In contrast to this approach we compute the path integral for the matrix model rather then just counting the states. The advantage of our approach is that it allows a conceptually simple (although technically evolved) extension to higher loops in the spirit of ``conventional'' perturbation theory. It appears that the description we obtain is in some sense complimentary to the one of the PET approach: for very long operators we get a $N^2$ scaling of the extensive thermodynamical functions, which signals for the \emph{string bit} phase, while for the operators below critical length the thermodynamical potentials scale as $N^0$, which is compatible with $N$-independent description obtained in the PET approach.

The plan of the paper is as follows. In the next section we review some basic properties of the matrix model: we rewrite it in the real form in terms of Hermitian matrices and pass to the ``second order formalism''. The second order formalism is important for the comparison the matrix model under study with the DLCQ-inspired BMN matrix model \cite{Berenstein:2002jq}. The comparison reveals that in our matrix model there are additional terms which seem unlikely to be canceled by the higher loop correction or to be generated from higher orders in the BMN matrix model.

Next, we describe the canonical quantization and conserved charges
of the model. In the third section we analyze the spin chains as
collective states of the matrix model. In particular, we analyze the
regimes at which the matrix model describes the gas of free spin
chains. In section \ref{sec:highT} we compute the free energy and
the entropy (which we will like to call `effective action') in the
high temperature limit. During the computation we observe a couple
of rather remarkable features of the model. One feature is that the
matrix model substantially simplifies in the mode expansion which
allows one to solve the model exactly at large $N$ analytically in
exponentials of the chemical potentials.

We analyze the situation in both cases of small and large chemical potentials and find out that for large enough value of the potentials (above the Hagedorn one) all terms scaling as $N^2$ and $N$ cancel out leaving us with the next leading contribution.\footnote{An approach based on random walks allows one to find this contribution too \cite{Sochichiu:2006yv}.} On the opposite side, when the chemical potentials are small the model behaves as a system of $N^2$ interacting particles.
In section \ref{sec:lowT} we consider small temperature limit. In this limit the typical behavior is one of a
$N$-particle system. Finally we give the string theory interpretation and discuss the results.

\newpage
% ----------------------------------------------------------------
\section{Matrix model}

\subsection{Dilatation operator and anomalous
dimensions of $\mathcal{N}=4$ SYM}

According to the AdS/CFT conjecture, the string states correspond to composite operators of the gauge theory. Furthermore, identification of the respective charges in the symmetry groups on both sides puts into correspondence the string energy levels to the eigenvalues of the dilatation operator of $\mathcal{N}=4$ SYM. Classically, the value of the dilatation operator is given by the composition of the dimensions of its elementary components. In the quantum theory this is corrected by the \emph{anomalous dimension} contribution. It also appears that the dilatation mixes different operators, which form, however, invariant classes which do not mix. A particular example which we analyze in this paper is given by the SU(2)-sector of SYM, which consists of all gauge invariant polynomial operators built from two complex scalars,
\begin{align}
  \Phi_1&\equiv Z=\ft12(\phi_5+\ii \phi_6),\\
  \Phi_2&\equiv \phi=\ft12(\phi_1+\ii \phi_2),
\end{align}
where $\phi_i$, $i=1,2,\dots,6$ are the scalars of the
$\mathcal{N}=4$ SYM.

The mixing matrix/dilatation operator in the context of the
perturbation theory was considered in
\cite{Beisert:2003tq,Beisert:2003jj}.

In particular, the one-loop contribution to the mixing matrix for operators in SU(2) sector was found to be \cite{Beisert:2002bb},
\begin{equation}
  H=-\frac{g_{\rm
    YM}^2}{16\pi^2}:\tr[\Phi^a,\Phi^b][\fc_a,\fc_b]:\equiv-\frac{g_{\rm
    YM}^2}{8\pi^2}:\tr[\phi,Z][\check{\phi},\check{Z}]:,
\end{equation}
where the checked character corresponds to the differential
operator,
\begin{equation}
   (\fc_a)_{mn}=\frac{\pd}{\pd\Phi^a_{nm}},
\end{equation}
and the colon ``:" denotes such an ordering in which no derivative
acts on the operators within the same group.

\subsection{The matrix action}
Consider the matrix model describing the anomalous dimensions of
$\mathcal{N}=4$ super Yang--Mills model. As it was found in
\cite{Bellucci:2004fh} this matrix model is described by the
action,\footnote{Note, however, the difference in the notations.}
\begin{multline}\label{act-cl}
  S(\Psi,\bar{\Psi},A)=\\
  \int \dd t\,\left(\tr
  \frac{\ii}{2}(\bar{\Psi}{}_a\nabla_0{\Psi}{}^a-\nabla_0
  {\bar{\Psi}}{}_a\Psi^a)-
  \tr\bar{\Psi}_a \Psi_a
  +\frac{g_{\rm YM}^2}{16\pi^2}
  \tr[\bar{\Psi}_a,\bar{\Psi}_b][\Psi^a,\Psi^b]\right)
\end{multline}
where\footnote{In \cite{Bellucci:2004fh} these fields are denoted by
$X$ and $\bar{X}$. Here, for convenience, we replace them by $\Psi$
and $\bar{\Psi}$, respectively. Also we introduce the term
$\tr\bar{\Psi}\Psi$ which describes the classical dimension.}
$\Psi_a$, $a=1,2$ are (non-Hermitian) $N\times N$ matrices, and
$\nabla_0 \Psi=\pd_0 \Psi+[A_0,\Psi]$, with Hermitian matrix $A_0$,
is the covariant time derivative.

\subsection{Real form}
To compare our model with other matrix models arising in the context
of AdS/CFT correspondence it is convenient to pass from the
variables $\Psi$ and $\bar{\Psi}$ to the Hermitian ones $X$ and $Y$
representing the Hermitian and the anti-Hermitian parts of $\Psi$ as
follows,
\begin{equation}\label{real}
  \Psi_a=\ft{1}{\sqrt{2}}(P_a+\ii X_a),
  \quad \bar{\Psi}_a=\ft{1}{\sqrt{2}}(P_a-\ii X_a).
\end{equation}
Substitution of \eqref{real} into the action \eqref{act-cl} yields,
\begin{multline}\label{act-real}
  S(P,X,A)=\\
  \int \dd t\,\left(\tr
  \ft{1}{2}\{P_a(\nabla_0X_a)-(\nabla_0P_a)X_a\}-
  \ft12(P_a^2+ X_a^2)\right.\\
  +\left.\ft{g_{\rm YM}^2}{64\pi^2}
  \tr\left\{[P_a,P_b]^2+[X_a,X_b]^2+2[P_a,X_b]^2\right\}\right),
\end{multline}
where we used the constraint $[P_a,X_a]=0$ and dropped off the total
time derivative term $(\ii/4)\ft{\dd}{\dd t}(P_a^2+X_a^2)$. The
action \eqref{act-real} can be written in a more economical way
using a common variable $X_A$ of extended dimensionality $A=1,\dots,
4$,
\begin{equation}
  X_A=(P_a,X_b).
\end{equation}
In this case the action \eqref{act-real} can be rewritten as
\begin{multline}\label{act-econ}
  S(X,A)=
  \int \dd t\,\left(
  \ft{1}{2}\omega_{AB}\tr X_A\nabla_0X_B-
  \ft12 \tr X_A^2
  +\ft{g_{\rm YM}^2}{64\pi^2}
  \tr[X_A,X_B]^2\right),
\end{multline}
where the anti-symmetric matrix $\omega_{AB}$ is given in terms of
$(X,Y)$ decomposition of $X_A$ by the following block structure
\begin{equation}
  \|\omega_{AB}\|=
  \begin{pmatrix}
    0 & \I\\
    -\I & 0
  \end{pmatrix}.
\end{equation}

Let us note that the form \eqref{act-econ} of the action is very
close to the one of the ``classical'' Yang--Mills type matrix
models. The only difference is the mass term and the first order
kinetic term.

\subsection{``Second order'' formalism}

The existend matrix models related to non-perturbative string dynamics
are second order acrtion i.e. their kinetic terms have the form of
velocity-squared. Our model is described by first order action
i.e. with at most linear velocity dependence. On the other hand the
matrix variables are non-Hermitian matrices, from which we can deduce
that the action \eqref{act-cl} the ``first order form'' of higher
order action. Indeed, the example of ordinary harmonic oscillator with
the classical action
\begin{equation}
  S=\int\dd t\,\left(\ft12\dot{x}^2-\ft12x^2\right),
\end{equation}
can be equivalently rewritten in the canonnical form
\begin{equation}
  S=\int\dd t\, \left(p\dot{x}-\ft12(p^2+x^2)\right),
\end{equation}
by the Legendre thransform with canonical momentum $p=\pd
\mathcal{L}/\pd\dot{x}$. Further one can pass to the complex
coordinate $a$ which is the classical counterpart of the oscillator
annihilation operator
\begin{equation}
  a=\ft1{\sqrt{2}}(p+\ii x),\quad
  \bar{a}=\ft1{\sqrt{2}}(p-\ii x).
\end{equation}
In the last coordinates the action takes the form
\begin{equation}
  S=\int\dd t
  \,\left(\ft12(\bar{a}\dot{a}-\dot{\bar{a}}a)-\bar{a}a\right)+\text{boundary
  terms}.
\end{equation}
If we replace $a$ with the matrix valued $\Psi$ and take the trace
we see that the quadratic part of the action \eqref{act-cl} is
nothing else that the harmonic matrix oscillator in the complex
first order form, while \eqref{act-real} gives the ordinary first
order formalism. Now, let us try to reconstruct the corresponding
``second order'' action. Classically, one can switch back to the
Lagrangian formalism just by solving equations of motion with
respect to $p$ in terms of $x$ and $\dot{x}$ and substituting $p$ in
the action by this solution. \footnote{In quantum theory in addition
to this one should
  also take care on modification of the measure (or of the scalar
  product).}

The equations of motions resulting from the variation of $P_a$ in
\eqref{act-real} read\footnote{We use the gauge $A=0$.}
\begin{equation}\label{p}
  P_a=\dot{X}_a-\ft{g_{\rm YM}^2}{16\pi^2}\left([P_b,[P_b,P_a]]
  +[X_b,[X_b,P_a]]\right).
\end{equation}
We can not solve this equation exactly, but we do not need the exact solution either. Recall that the action \eqref{act-cl} is only the order $g_{\rm YM}^2$ approximation of the ``exact'' matrix model describing the anomalous dimensions to all orders in perturbation theory as well as nonperturbative effects. So, for us it suffices to solve the equation \eqref{p} to only the first order in $g_{\rm
  YM}^2$. This solution is given by,
\begin{equation}
  P_a=\dot{X}_a-\ft{g_{\rm
  YM}^2}{16\pi^2}\left([\dot{X}_b,[\dot{X}_b,\dot{X}_a]]
  +[X_b,[X_b,\dot{X}_a]]\right).
\end{equation}

Then, the resulting ``second order formalism'' action takes the
following form (up to the order $g_{\rm YM}^2$),
\begin{multline}\label{second-order}
  S=\\
  \int\dd t\, \tr\left(\ft12 (\dot{X}^2- X^2)+
  \ft{g_{\rm YM}^2}{64\pi^2}\left([X_a,X_b]^2+2[\dot{X}_a, X_b]
  +[\dot{X}_a,\dot{X}_b]^2\right)
  \right).
\end{multline}
The Gauss law constraint at the same time takes the following form,
\begin{equation}
  G=[\dot{X}_a, X_a]+\ft{g_{\rm YM}^2}{16\pi^2}
  \left([\dot{X}_b,[\dot{X}_b,\dot{X}_a]]
  +[X_b,[X_b,\dot{X}_a]]\right).
\end{equation}

Let us note the fact, that for slow varying matrices: $\dot{X}\to 0$, the matrix action \eqref{second-order} is equivalent to the matrix model of \cite{Berenstein:2005jq}. This is not surprising since the respective matrix model reproduce the BMN spectrum, while our model reproduces the leading order in $\lambda$ of this spectrum. The surprising part is that beyond the ordinary commutator term our action contains also commutator of velocities. One can see that such terms can not be canceled by higher loop contribution. It would be interesting to see wether these terms can be removed by a proper redefinition of the fields.

\subsection{Quantization, gauge invariance }

To quantize the model given by the action \eqref{act-cl} we have to
impose first the gauge fixing condition. The most natural condition in
our case is the temporal gauge $A=0$.

As it can be seen, the variation of the action with the respect to
the gauge field $A$ fails to produce a \emph{bona fide} equations of
motion, one has instead the Gauss law constraint,
\begin{equation}\label{gl}
  G=[\Psi_a,\bar{\Psi}_a]=\ft12\omega_{AB}[X^A,X^B]\approx 0.
\end{equation}

From the simplectic structure of the classical action \eqref{act-cl}
one can extract the canonical Poisson bracket of the model. In
components it reads:
\begin{equation}\label{pb-c}
  \{(\bar{\Psi}_a)_{i}{}^{j},(\Psi_b)_{k}{}^l\}=-\ii\delta_{i}{}^{l}
\delta_k{}^j.
\end{equation}
This can be written in the componentless form using the ``quantum
group'' notations
\begin{equation}\label{pb-inv}
  \{\bar{\Psi}_a^{(1)},\Psi_b^{(2)}\}=-\ii P_{12},
\end{equation}
where the $N^2\times N^2$-dimensional matrices $\Psi_a^{(1,2)}$ and
$P_{12}$ are defined
\begin{equation}
  \Psi_a^{(1)}=\Psi_a\otimes \I,\quad \Psi_a^{(1)}=\I\otimes \Psi_a,
\end{equation}
while $P_{12}$ is the permutation operator
\begin{equation}
  P_{12}\,a\otimes b=b\otimes a.
\end{equation}

The canonical quantization consist in the promoting of the Poisson
bracket in either form \eqref{pb-c} or \eqref{pb-inv} to the quantum
commutator for which
\begin{equation}\label{qa}
  \qp{(\bar{\Psi}_a)_{i}{}^{j},(\Psi_b)_{k}{}^l}=
  \delta_{i}{}^{l}\delta_k{}^j .
\end{equation}
The interpretation of the quantum algebra \eqref{qa} is the following. The operator $(\bar{\Psi}_a)_i{}^j$ creates a matrix element in the row $i$ and column $j$ having the polarization $a$ while the respective component of $(\Psi_a)_j{}î$ destroys it. Let us note that the quantum commutator is denoted by the double braces in order to distinguish it from the ordinary matrix one which is related to different contractions of the matrix indices $i,j,\dots$ but not related to the permutation of the operator valued matrix elements.

In the complete analogy with the Harmonic oscillator the (extended) Fock space can be constructed by the action of the rising operators $(\bar{\Psi}_a)_i{}^j$ on the oscillator vacuum state
\begin{equation}\label{fock}
  \ket{C}=\sum_{k} (C^{a_1\dots a_k})_{j_1\dots j_k}^{i_1\dots i_k}
  (\bar{\Psi}_{a_1})_{i_1}{}^{j_1}\dots (\bar{\Psi}_{a_k})_{i_1}{}^{j_k}
  \ket{\Omega},
\end{equation}
where
\begin{equation}
  (\Psi_a)_{i}{}^{j}\ket{\Omega}=0,\quad \forall a,i,j.
\end{equation}

Due to the constraint \eqref{gl} the extended Fock space is too large. One has to restrict it to the gauge invariant subspace by imposing the condition,
\begin{equation}\label{phys}
  \hat{G}\ket{\Phi}=0,
\end{equation}
where $\hat{G}=:[\Psi_a,\bar{\Psi}_a]:$ is the quantum version of the constraint \eqref{gl}. (Where the colon denotes an ordering prescription according to which all the components of $\bar{\Psi}$ stay always to the right of those of $\Psi$.)

As in the case of ordinary gauge fields the operator $\hat{G}$
corresponding to the constraint is the generator of gauge
transformations:
\begin{equation}
  \qp{\hat{G}(u),\Psi_a}=-[u,\Psi_a], \quad
\qp{\hat{G}(u),\bar{\Psi}_a}=[u,\bar{\Psi}_a],
\end{equation}
where $\hat{G}(u)$ is the operator $G$ smeared with an element of
su(N) algebra $u$:
\begin{equation}
  \hat{G}(u)=\tr \hat{G}u=\tr :[\bar{\Psi}_a,\Psi_a]:u.
\end{equation}

As a result of application the condition \eqref{phys} the physical
space is generated by only those $\ket{C}$ which correspond to gauge
invariant polynomials of $\Psi_a$. In terms of equation \eqref{fock}
this means that the quotients $(C^{a_1\dots a_k})_{j_1\dots
 j_k}^{i_1\dots i_k}$ should be linear combinations of the products of
delta symbols $\delta_{i}{}^j$. The most generic such monomial of
order $L$ is given by a permutation of $L$ elements $\gamma\in S_L$,
\begin{equation}
  (C^{a_1\dots a_k})_{j_1\dots j_k}^{i_1\dots i_k}=
  C^{a_1\dots a_k,\gamma}
  \delta_{j_1}{}^{i_{\gamma_1}}\dots \delta_{j_k}{}^{i_{\gamma_k}},
\end{equation}
where $\gamma_k=\gamma(k)$ is the permutation of the $k$-th element.

\subsection{Conserved charges}
In the model \eqref{act-cl} there is a number of quantities which
are conserved. Among these let us consider the quadratic ones
\begin{equation}
  L_{ab}=\tr \bar{\Psi}_a\Psi_b , \quad L_{ab}^\dag=L_{ba},\quad
  \dot{L}_{ab}=0.
\end{equation}
They can be split into Hermitian operator $L$ and operator valued
vector $\vec{S}$,
\begin{equation}\label{l-and-s}
  L=L_{aa}, \quad \vec{S}=\vec{\sigma}_{ab}L_{ab},
\end{equation}
where $\vec{\sigma}$ is the vector of Pauli matrices. As we will see below these operators correspond to the total length and total spin operators of the spin chains. Both $L$ and $\vec{S}$ generate a representation of the algebra U(2) which can be split into irreducible components described by positive integer value of $L$ (U(1) spin) and the spin $s$ irreducible representation of the SU(2) component.

In particular, due to the normal ordering of the operators all above
charges vanish on the vacuum state $\ket{\Omega}$. The value of the
operator $L$ corresponds to the total number of the excited oscillator
modes, while e.g. $S_3=\ft12\tr \Psi_1\bar{\Psi}_1-\Psi_2\bar{\Psi}_2$
corresponds to the difference in the number of excited modes for the
first and the second oscillator.

\subsection{Matrices as spin chain gas}
As we discussed earlier, the quantum Hamiltonian of the matrix model can be mapped to the Hamiltonian of a spin system with a chaining interaction \cite{Bellucci:2004ru,Bellucci:2004qx} (see also \cite{Agarwal:2004cb,Agarwal:2004sz}). The limit $N\to\infty$ with $\lambda=g^2 N$-fixed reduces that model to the integrable Heisenberg XXX$_{1/2}$ model.
In this picture spin chains correspond to one-trace operators. As soon as $N$ is finite the subspace of such operators is not invariant with respect to dynamics and, therefore, in a gauge invariant configurations one is forced to consider a mixture of matrix states with all possible number of traces.

Naively, in $N\to\infty$ limit the non-planar interaction vanishes and one ends up with an ``ideal gas'' of spin chains, where each chain is conserved by the dynamics.

This simple picture, which we have for the spin chain/matrix model correspondence in the quantum theory, breaks down as soon as we consider the (semi)classical limit of either spin chains or matrix theory. As it turns out, the semiclassical regimes for two descriptions appear at different if not contradictory conditions. Indeed, the matrix model when the expectation values for the occupation numbers of most modes are larger than one. This gives us the condition,
\begin{equation}\label{evlN}
  L=\tr \bar{\Psi}_a\Psi_a\sim N^2.
\end{equation}
On the other hand, as we know from the BMN evaluation, which should hold also in our case, the non-planar interaction rate is given by a factor proportional to $L^2/N$. Then, the regime \eqref{evlN} for the spin description means that the non-planar interactions are not only strong but overwhelmingly dominating the dynamics. In this case one can not even speak about the spin chains. In contrast, for $L\ll\sqrt{N}$, when one can neglect the non-planar interaction and, hence, the spin chains are stable the matrix model is in essentially quantum regime.

A gauge invariant matrix state generally will contain a mixture (gas, condensate, etc.) of spin chains of different lengths. The dynamics of this ``spin soup'' is such that the very short chain will tend to join bigger ones while too long chains will tend to split into smaller ones. Therefore, an equilibrium state should be statistically dominated by chains of a particular length.

As one can see, the statistical description of such a complicate system appears very naturally as it can catch the properties of the dynamics of both spins and matrices. Moreover, moving from one regime to another we can extract the relevant information like which sort of behavior dominates at different regimes. We do this in the next sections.

\section{Spin chain gas at high temperature}\label{sec:highT}

Let us consider the closed sector of matrix model given by states of the length  $L$, and let us further restrict ourself to the states of definite value of spin $\vec{S}$. The partition function restricted to such a subspace of the Hilbert space is given by,
\begin{equation}
  Z(L,\vec{S})=\tr\Pi(L,\vec{S})\e^{-\beta H(\bar{\Psi},\Psi)},
\end{equation}
where $\Pi(L,\vec{S})$ is the projector to the
$(L,\vec{S})$-subspace and $H(\bar{\Psi},\Psi)$ is the Hamiltonian
operator of \eqref{act-cl}. The projection to given values of $L$
and $\vec{S}$ can be realized in the following way:
\begin{equation}\label{partial}
  Z(L,\vec{S})=\int\dd\mu\dd\vec{x}\,\e^{\mu L+\vec{x}\cdot
  \vec{S}}Z(\mu,\vec{x}),
\end{equation}
where $Z(\mu,\vec{x})$ is the (grand canonical) partition function:
\begin{equation}\label{gc-pf}
  Z(\mu,\vec{x})=\tr \e^{-H_{\beta,\mu,\vec{x}}(\bar{\Psi},\Psi)},
\end{equation}
where the chemical potentials $\mu$ and $\vec{x}$ enter into the
Hamiltonian $H_{\mu,\vec{x}}(\bar{\Psi},\Psi)$ in the following way,
\begin{multline}\label{ham}
  H_{\beta,\mu,\vec{x}}(\bar{\Psi},\Psi)=\\
  \tr\left(-\ii\bar{\Psi}_a[A,\Psi_a]+\mu\bar{\Psi}_a \Psi_a+\ft12
  \vec{x}\bar{\Psi}_a\vec{\sigma}_{ab} \Psi_b
  -\frac{\beta g_{\rm YM}^2}{16\pi^2}[\bar{\Psi}_a,
  \bar{\Psi}_b][\Psi^a,\Psi^b]\right).
\end{multline}
The original Hamiltonian is recovered at $\mu=\beta$ and
$\vec{x}=0$.

To find the partial partition function \eqref{partial} let us first
compute the grand canonical partition function $Z(\mu,\vec{x})$.

An important issue in this computation is one of gauge symmetry and
the gauge fixing. There are several possibilities with this. One
possible approach is given by imposing the diagonal gauge fixing to
one of the components, say the real part of $\Psi_1$. As is claimed
in \cite{Berenstein:2005jq} this leads to the description of the
model in terms of a fermionic liquid. In the present case we use a
different approach. To discuss the problem of gauge fixing it is
useful to switch for a while to the path integral interpretation of
the partition function. In this interpretation the inverse time
$\beta$ plays the role of the size of compact Euclidean time.
Therefore, the path integral formulation for the partition function
implies periodic boundary conditions for the matrices:
$X(\tau+\beta)=X(\tau)$.

Let us now turn to the gauge field. It is clear, that because of periodic time we can not impose the temporal gauge: $A=0$. Indeed, out of the gauge field $A(\tau)$ one can form a gauge covariant object called Polyakov loop\footnote{This equation defines $W[A]$ up to a gauge transformation related to the choice of initial point.}
\begin{equation}
  W[A]=T\exp\oint\dd\tau\, A(\tau),
\end{equation}
whose eigenvalues are gauge invariant and, therefore, can not be canceled by a gauge transformation. Hence, the admissible gauge we can impose is to fix $A$ to be at most constant and diagonal.

In the operator formalism one can still overcome the above restrictions and impose the temporal gauge $A=0$, but one has supply it with the Gauss law constraint on the matrices. As the Gauss law constraint is a conserved quantity, it suffices to impose it once at a preferred time instant. This is equivalent to the restriction of the gauge field $A$ in the Hamiltonian \eqref{ham} to be constant and diagonal.

The Gauss law constraint act on the quantum states as generator of gauge transformations. Hence, vanishing of the constraint is equivalent to the gauge invariance of the respective state. As we discuss in the Appendix \ref{app:gauge-inv}, in the case of compact Lie gauge group, projecting to the subspace of gauge invariant states can be implemented by the group integration. The advantage of this approach is that the compactness of the gauge group is taken into account as well.

At large temperature and fixed value of the length operator $L$ the commutator term of the Hamiltonian \eqref{ham} is suppressed by a factor $\beta g_{\rm YM}^2$, therefore we can treat it as a perturbation over the quadratic part of the Hamiltonian.

In what follows we will use the holomorphic representation which corresponds to the quantum oscillator described by the quadratic part of the Hamiltonian \eqref{ham}. For the convenience of the reader we give a brief summary of the holomorphic representation in the Appendices \ref{app:holomorphic} and \ref{app:gauge-inv}.

\subsection{Quadratic part}\label{sec:quadratic}

As we discussed above, we can consider the commutator term to be a
small perturbation over the quadratic part of the Hamiltonian, such
that we can apply the perturbation theory. Let us first compute the
bare partition function which is the trace over gauge invariant
subspace,
\begin{equation}
  Z_0(\beta)=\tr_{g.i.}\e^{- H_{0,\mu,\vec{x}}},
\end{equation}
where $H_{0,\mu,\vec{x}}$ is the quadratic part of the Hamiltonian
\eqref{ham}, which consists of chemical potentials only:
\begin{equation}\label{ham0}
  H_{0,\mu,\vec{x}}(\bar{\Psi},\Psi)=
  \tr\left(\mu\bar{\Psi}_a
  \Psi_a+\ft1 2
  \bar{\Psi}_a(\vec{x}\cdot\vec{\sigma})_{ab} \Psi_b
  \right).
\end{equation}
The Gaussian integral in \eqref{ham0} is convergent provided
$\sqrt{\vec{x}^2}<\mu$. For lager values of $\vec{x}$ it diverges
because the quadratic form in \eqref{ham0} becomes indefinite.
Physically, however, more interesting situations are when spin is
suppressed stronger than the length operator. In principle this can
be done by a ``Wick rotation'' to imaginary spin chemical potential,
$\vec{x}\mapsto \ii\vec{x}$.

Restriction of the trace to the gauge invariant subspace can
implemented, as discussed in Appendix \ref{app:gauge-inv}, by
averaging over the \emph{global} gauge group,
\begin{multline}\label{gi-tr}
  Z_0(\mu,\vec{x})\equiv\tr_{g.i.}\e^{- H_{0,\mu,\vec{x}}}=\\
  \int\dd U\int\frac{\dd\bar{\Psi}\dd
  \Psi}{\pi^{N^2}}\,\exp\left(-\tr\bar{\Psi}_aU^{-1}\Psi_a U+\bar{\Psi}
  \e^{-\mu-\vec{x}\cdot \vec{\sigma}/2}\Psi\right),
\end{multline}
where the exponent in the last term is understood as a $2\times 2$
matrix exponent. Let us note that, in contrast to ordinary gauge
fixing, averaging over the gauge group keeps the gauge invariance
explicit i.e. the integral is invariant with respect to global gauge
transformations:
\begin{equation}
  U\to V^{-1}UV,\quad \Psi\to V^{-1}\Psi,\quad \bar{\Psi}\to
  V^{-1}\bar{\Psi}V.
\end{equation}

In the basis of eigenvectors $\ket{\pm}$ of the matrix $\vec{x}\cdot
\vec{\sigma}$,
\begin{equation}
  \vec{x}\cdot \vec{\sigma}\ket{\pm}=\pm x\ket{\pm},\qquad
  x=\sqrt{\vec{x}^2},
\end{equation}
the matrix $\e^{-\mu-\vec{x}\cdot \vec{\sigma}/2}$ becomes diagonal
with values $\e^{-\mu_{\pm}}$, where
\begin{equation}
  \mu_{\pm}=\mu\pm\frac{x}{2}.
\end{equation}

The symmetry which is left allows us fixing the matrix $U$ to be diagonal.\footnote{This is in the direct relation to the fact that the gauge field can also be fixed to be constant diagonal.} Also changing the variable in the $U$-integral to the diagonal values $U=\diag \{\e^{\ii \theta_n}\}$ produces square of the Vandermonde determinant,
\begin{equation}\label{vandermond}
  |\Delta(\theta)|^2=\prod_{m>n}2(1-\cos\theta_{mn}),\qquad \theta_{mn}=\theta_m-\theta_n.
\end{equation}
We also used the fact that the integral does not depend on the U(1) factor. Hence one can fix the sum of the eigenvalues (`center of mass') to vanish,\footnote{In what follows we just systematically drop the contributions proportional to $\sum\theta_n$.}
\begin{equation}
  \sum_{n}\theta_n=0\mod 2\pi.
\end{equation}

As a result of above transformations the gauge invariant trace takes
the following `normal form',
\begin{multline}\label{h-norm}
  Z_0(\mu,\vec{x})=\int\prod_n\dd\theta_n\,|\Delta(\theta)|^2\\
  \times\int \frac{\dd \bar{\Psi}\dd\Psi}{\pi^{N^2}}
  \exp\left(-\sum_{\substack{m,n\\
  \pm}}h^{(\pm)}_{(mn)}(\mu,\vec{x},\theta)\bar{\Psi}^\pm_{mn}\Psi^\pm_{nm}\right),
\end{multline}
where the normal modes $h^{(\pm)}_{(mn)}(\mu,\vec{x},\theta)$, are
given by,
\begin{equation}\label{h-pm}
  h^{(\pm)}_{mn}=\e^{\ii\theta_{mn}}-\e^{-\mu_{\pm}}.
\end{equation}

As one can see, the integral over the matrices $\Psi$ and
$\bar{\Psi}$ is Gaussian and can be easily taken. The quadratic
partition function then is reduced to the integral over $\theta$'s
only,
\begin{multline}\label{z0alpha}
  Z_0(\mu,\vec{x})=\int\prod_n\dd\theta_n\,|\Delta(\theta)|^2
  \prod_{m,n}\left[h^{(+)}_{mn}h^{(+)}_{nm}h^{(-)}_{mn}h^{(-)}_{nm}
  \right]^{-1}\\
  =\frac{2^{-\ft12N(N+1)}\e^{N^2\mu}}{[\sinh(\mu_{+}/2)\sinh(\mu_{-}/2)]^{N}}
  \int\prod_n\dd\theta_n\,\times\\\prod_{m>n}
  \frac{1-\cos\theta_{mn}}
  {(\cosh\mu_{+}-\cos\theta_{mn})(\cosh\mu_{-}-\cos\theta_{mn})}
  .
\end{multline}

Using \eqref{partial}, one can pass to the variables $L$ and $\vec{S}$. As soon as a large number of degrees of freedom is concerned this integration can be done by the saddle-point approximation, i.e. by replacing $\mu$ and $\vec{x}$ with their solutions in the presence of the sources $L$ and $\vec{S}$ to the equations of motion corresponding to the effective action resulting from \eqref{z0alpha}. The saddle point conditions coincide with the Legendre equations. Hence the Laplace transform in this limit is reduced to the Legendre transform of the free energy:
\begin{subequations}\label{extr}
\begin{align}\label{LS}
  &L=\frac{\pd F(\mu,\vec{x})}{\pd\mu},\qquad
  \vec{S}=-\frac{\pd F(\mu,\vec{x})}{\pd\vec{x}};\\
  \label{legendre}
  &S_{\mathrm{eff}}(L,\vec{S})=L\mu(L,\vec{S})+\ii\vec{S}\cdot \vec{x}(L,\vec{S})
  -F(\mu(L,\vec{S}),\vec{x}(L,\vec{S}))
\end{align}
\end{subequations}
where functions $\mu(L,\vec{S})$ and $\vec{x}(L,\vec{S})$ in the
second line are found as solutions to the first line equations
\eqref{LS}. The free energy $F$ is defined as the log of the
partition function,
\begin{equation}\label{s-eff}
  F(\mu,\vec{x})=-\ln Z_0(\mu,\vec{x}).
\end{equation}
Thus, to obtain the effective action we have to integrate over the gauge field eigenvalues $\theta_n$. In some cases this can be \emph{also} done by the saddle-point approximation, i.e. replace the integral by the value of the integrand at its maxima.

\subsection{The free energy}

Consider the saddle point approach to the integral \eqref{z0alpha}. Finding the maximal value of the integrand in the partition function \eqref{z0alpha} is reduced to the problem for finding extrema with respect to variation of $\theta_n$ for the following action,
\begin{multline}\label{seff-alpha}
  F(\theta;\mu,\vec{x})=-N^2\mu+N[\ln\sinh(\mu_{+}/2)+\ln\sinh(\mu_{-}/2)]\\
  +\ft12\sum_{\substack{m,n\\
  n\neq  m}}\left(-\ln(1-\cos\theta_{mn})+\ln(\cosh\mu_{+}-\cos\theta_{mn})
  +\ln(\cosh\mu_{-}-\cos\theta_{mn})\right),
\end{multline}
where we dropped the constant terms, which can be absorbed by a
redefinition of the measure.

The equations for the Legendre transform \eqref{extr} then read,
\begin{subequations}\label{leg}
\begin{align}\nonumber
  L&=-N^2+\ft12N[\coth(\mu_{+}/2)+\coth(\mu_{-}/2)]\\ \label{leg-nl}
  &+\ft12\sum_{\substack{m,n\\
  n\neq  m}}\left(\frac{\sinh \mu_{+}}{\cosh\mu_{+}-\cos\theta_{mn}}
  +\frac{\sinh \mu_{-}}{\cosh\mu_{-}-\cos\theta_{mn}}\right)\\
  \nonumber
  2\vec{S}&=\ft12N[\coth(\mu_{+}/2)-\coth(\mu_{-}/2)]\\
  &+\ft 12\sum_{\substack{m,n\\
  n\neq  m}}\left(\frac{\sinh \mu_{+}}{\cosh\mu_{+}-\cos\theta_{mn}}
  -\frac{\sinh
  \mu_{-}}{\cosh\mu_{-}-\cos\theta_{mn}}\right)\frac{\vec{x}}{x},\\
  \nonumber
   \sum_{m}&\left(-\frac{\sin\theta_{nm}}{1-\cos\theta_{nm}}+
   \frac{\sin\theta_{nm}}{\cosh\mu_{+}-\cos\theta_{nm}}+
   \frac{\sin\theta_{nm}}{\cosh\mu_{-}-\cos\theta_{nm}}\right)\\ \label{eom-alpha-nl}
  &\qquad =0,
\end{align}
\end{subequations}
where the summation in the last equation should run through $m$
different from $n$. As some $\theta_m$ approaches the $\theta_n$,
the first term in \eqref{eom-alpha-nl} diverges as $\sim
1/(\theta_n-\theta_m)$. In what follows we assume that this term is
regularized by replacing $1$ by by a term $\cosh\varepsilon$, where
$\varepsilon$ afterwards will be sent to zero. Having this
regularization in mind we can extend the last sum to all $m$.

Let us first discuss the equations \eqref{leg} from the qualitative point of view. The last equation, \eqref{eom-alpha-nl} describes an equilibrium static configuration of the system consisting of $N$ particles on a circle with the following pairwise interaction potential,
\begin{equation}
  \varphi=-\ft12\ln(1-\cos\lambda)+\ft12\ln(\cosh\mu_{+}-\cos\lambda)+
  \ft12\ln(\cosh\mu_{-}-\cos\lambda).
\end{equation}

As one can see, when $\mu_{\pm}\neq 0$ the interaction potential consists of one repulsive and two attracting terms. Normally, at small distances the repulsive term dominates while at large separations the interaction becomes attractive. Now several variants are possible. First, suppose the scale at which the attractive force starts to win over the repulsive one is larger than $2\pi$. In this case the repulsive dominant interaction will force the eigenvalues to arrange \emph{uniformly} over the circle. This is a configuration with unbroken subgroup of the translation symmetry for $\theta$: $\theta_n\mapsto \theta_{n}+\eta$. This happens when $\mu_{\pm}$ is large enough. When $\mu_{\pm}$ decrease the scale at which interaction becomes attractive decreases as well and at some point it becomes less than the circle's length $2\pi$. At this stage for eigenvalues it becomes more ``convenient'', from the point of view of ``total energy'', to condense around some point(s) rather than be uniformly spread over the circle. In particular, when at least one of $\mu_{\pm}$ vanishes the repulsion term is canceled completely and all eigenvalues tend to collapse to a single point. The last is the extremal case of breaking of the translational symmetry of the eigenvalue distribution.

\subsubsection{Expansion in powers of $\e^{-\mu_{\pm}}$}
Let us pass to a quantitative analysis rather a qualitative description. The problem dramatically simplifies if one looks for solutions which are analytic in $\e^{-\mu_{\pm}}$. This appears possible due to the fact that the expansion in terms of powers of $\e^{-\mu_{\pm}}$ of the `equation of motion' \eqref{leg} is remarkably simple and takes the following form,
\begin{equation}\label{eom-alpha-l-large}
  2\sum_{\omega=1}^{\infty}\left(-1+\e^{-\omega\mu_{+}}+\e^{-\omega\mu_{-}}\right)
  \sum_{\substack{m\\
  }}
  \sin(\omega\theta_{nm})=0.
\end{equation}
As one can see, this expansion organizes itself into something
similar to a Fourier series.

Applied to the free energy \eqref{seff-alpha} the expansion yields,
\begin{multline}\label{seff-alpha-expnd}
  F(\theta;\mu,\vec{x})=
  N[\ln\sinh(\mu_{+}/2)+\ln\sinh(\mu_{-}/2)-\mu]\\
  +\sum_{\omega=1}^{\infty}\frac{1}{\omega}(1-
  \e^{-\omega \mu_{+}}-\e^{-\omega \mu_{-}})\sum_{\substack{m,n\\
  m\neq  n}}\cos(\omega\theta_{mn}).
\end{multline}
In \eqref{eom-alpha-l-large} and \eqref{seff-alpha-expnd} one can
observe that the terms scaling as $N^2$ were canceled away by the
zeroth term in the expansion. To obtain \eqref{eom-alpha-l-large}
and \eqref{seff-alpha-expnd} we used the following expansion,
\begin{equation}
  \ln\left(\cosh\mu-\cos\theta\right)=
  \mu-\ln 2-\sum_{n=1}^{\infty}\frac{\e^{-n\mu}}{n}\cos[n\theta],
\end{equation}
as well as its derivative.

Let us continue with the succession of cancelations. Observe the
following properties of trigonometric sums,
\begin{subequations}\label{sin-cos}
\begin{align}
  \sum_{m}\sin\omega\theta_{mn}&=\sin\omega\theta_n\sum_{m}\cos\omega\theta_m
  -\cos\omega\theta_n\sum_{m}\sin\omega\theta_m,\\
  \sum_{\substack{m,n\\
  m\neq
  n}}\cos\omega\theta_{mn}&=\left(\sum_{m}\cos\omega\theta_m\right)^2+
  \left(\sum_{m}\sin\omega\theta_m\right)^2\\ \nonumber
  &\qquad-\sum_{m}\left(\cos^2\omega\theta_m+\sin^2\omega\theta_m\right)=
  \tilde{\rho}_{\omega}^2+\hat{\rho}_{\omega}^2-N,
\end{align}
\end{subequations}
where,
\begin{equation}\label{rho-fourier}
\tilde{\rho}_\omega=\sum_n\cos(\omega\theta_n),\quad
\hat{\rho}_\omega=\sum_n\sin(\omega\theta_n)
\end{equation}
for $\omega=1,2,3,\dots$

It is natural to assume that the equilibrium $\theta$-distribution
possesses at least one reflection symmetry such that one can choose
a point with respect to which the distribution of $\theta$'s is
\emph{even}. In this case the sums over sines $\hat{\rho}_{\omega}$
vanish and we are left with the following equations
\begin{align}\label{eom-alpha-exp}
  \sum_{\omega=1}^{\infty}&\left(-1+2\e^{-\omega\mu}\cosh\left(\frac{\omega x}{2}\right)\right)
  \sin(\omega\theta_n)\tilde{\rho}_{\omega}=0,
\end{align}
as well as with
\begin{equation}\label{free-exp}
  F(\theta;\mu,\vec{x})=\sum_{\omega=1}^{\infty}\frac{1}{\omega}\left(1-
  2\e^{-\omega\mu}\cosh\left(\frac{\omega x}{2}\right)\right)\tilde{\rho}_{\omega}^2.
\end{equation}
A glance at \eqref{eom-alpha-exp} and \eqref{free-exp} reveals that
the free terms scaling linearly in $N$ are canceled if we extend
summation to all possible pairs of eigenvalue numbers.

It is interesting to note that equations of motion do not depend on
the constant part of eigenvalue distribution $\tilde{\rho}_0$.
Hence, an obvious solution to the equation \eqref{eom-alpha-exp} is
given by the uniform distribution of the eigenvalues: $\theta_n=2\pi
n/N$, $n=0,1,2\dots,N-1$, or $\rho(\lambda)=N/2\pi\equiv \constant$.

Let us find the other solutions. In the limit of large number of
eigenvalues their distribution can be described by a continuous
density $\rho(\theta)$ such that,
\begin{equation}
  \int_{-\pi}^{\pi}\dd\theta\,\rho(\theta)=N,\qquad \rho(\theta)\geq 0,
\end{equation}
and a sum over eigenvalues can be replaced by an integral as
follows,
\begin{equation}
  \sum_nf(\theta_n)
  =\oint\dd\theta\,\rho(\theta)f(\theta),
\end{equation}
for some function $f(\theta)$ which is smooth enough. In particular,
$\tilde{\rho}_{\omega}$ in this limit become coefficients of Fourier
series,
\begin{equation}
  \tilde{\rho}_\omega=\oint\dd\lambda\,\rho(\lambda)\cos(\omega
  \lambda), \quad
  \rho(\lambda)=\rho_0+\sum_{\omega=1}^\infty
  \frac{\tilde{\rho}_{\omega}}{\pi}\cos\omega\lambda.
\end{equation}

Consider now the equation \eqref{eom-alpha-exp}. The equality in
\eqref{eom-alpha-exp} holds when each term in the sum vanishes
separately i.e.,
\begin{equation}\label{eom-mode}
  \tilde{\rho}_{\omega}
  \left(-1+2\e^{-\omega\mu}\cosh\left(\frac{\omega x}{2}\right)\right)=0,\qquad
  \omega=1,2,3\dots
  .
\end{equation}

Eq. \eqref{eom-mode} in particular means that the Fourier expansion
of $\rho(\theta)$ can have only such modes
($\tilde{\rho}_{\omega}\neq 0$) for which the expression inside the
parenthesis of \eqref{eom-mode} vanishes by itself,
\begin{equation}\label{irr}
  -1+2\e^{-\omega\mu}\cosh\left(\ft{\omega x}{2}\right)=0, \qquad
  \omega\in \mathbb{Z}_{+}.
\end{equation}

So, let us analyze the possible solutions to \eqref{irr}. Basically
the equation \eqref{irr} admits at most one solution when
$\mu\geq\ft x2$. Even in this case, due to the discrete nature of
$\omega$ such a solution is possible only for particular pairs
$(\mu,x)$. However, when $\mu$ and $x$ become small $\omega$ can be
treated as a continuous variable, therefore the set of pairs
$(\mu,x)$ for which one can find an integer solution becomes dense.

In general, for each value of $\omega$ there is a one-parameter
family of $(\mu_{\omega},x_{\omega})$ satisfying \eqref{irr}.

Now, let us turn to the ``free energy'' $F(\mu,\vec{x})$. Due to the
factor $(1-2\e^{-\omega\mu}\cosh\ft{\omega x}{2})$ multiplied to
each mode $\tilde{\rho}_{\omega}$ the ``on-shell'' value of
$F(\theta;\mu,\vec{x})$ vanishes. Physically this can be related to
the following fact. For simplicity consider the well-studied case of
the gauged one-matrix oscillator. This system, unlike the
multi-matrix oscillator, can be solved by imposing the diagonal
gauge to the matrix field. The Faddeev--Popov determinant, or better
to say the Jacobian, which arises when passing to the description in
terms of eigenvalues is again a Vandermonde determinant. Upon
quantization the determinant can be absorbed into the wave function
by rescaling it by the square root of the Vandermonde determinant.
Such a rescaling makes the wave function odd with respect to
permutation of the eigenvalues of the matrix field. Thus, the
diagonal matrix oscillators behave like fermionic ones. In
particular, they have to obey the Pauli exclusion principle: no two
diagonal oscillators could be in the same state. One of the effects
of this is that the states with $L<N$ are banned. In terms of the
chemical potential this means that for $\mu$ larger than a critical
value $\mu_{c}$ the free energy should become trivial i.e.
independent of $\mu$. Rather than disappointing this fact seems to
be compatible with the PET calculation of the partition function
\cite{Spradlin:2004pp}, which establishes that the free energy in
this regime is finite (and $N$ independent) as $N$ goes to infinity.

In spite that it appears that our approximation was not sensitive enough to obtain some nonzero thermodynamical quantities we still can try to extract some information from the system by observing that writing down the Legendre transformation equations \eqref{leg} before solving equations of motions for $\theta$'s lead to a non-trivial result for $L$ and $S$. This corresponds to the interchange of integration in $\theta$'s with one of $\mu$ and $x$.

In the case of a single nontrivial mode $\omega$, satisfying
\eqref{irr}, the Legendre equations \eqref{leg} are reduced to,
\begin{subequations}\label{leg-fin}
\begin{align}
  L&=\tilde{\rho}_{\omega}^2,\\
  2S&=\tanh\left(\frac{\omega x}{2}\right)\tilde{\rho}_{\omega}^2,
\end{align}
\end{subequations}
where we used \eqref{irr} to eliminate the factor $2\e^{-\omega
\mu}\cosh(\omega x/2)$ in the first equation and $2\e^{-\omega\mu}$
in the second one. Equations \eqref{leg-fin} together with the mode
equation \eqref{irr} seem to be enough to eliminate $\mu$ and $x$ in
favor of $L$ and $S$ for given $\omega$:
\begin{subequations}\label{leg-sol}
\begin{align}
  x&=\frac{2}{\omega}\tanh^{-1} \left(\frac{2S}{L}\right),\\
  \mu&=\frac{1}{\omega}\ln\frac{2}{\sqrt{1-\left(\frac{2S}{L}\right)^2}}.
\end{align}
\end{subequations}

As the ``free energy'' $F$ vanishes on-shell the effective action
$S_{\rm eff}(L,\vec{S})$ is given by,
\begin{multline}\label{noN}
  S_{\omega}=L\mu(L,\vec{S})+\vec{S}\cdot x(L,\vec{S})=\\
  \frac{1}{\omega}\left(L\ln\frac{2}{\sqrt{1-\left(\frac{2S}{L}\right)^2}}+
  2S\tanh^{-1} \left(\frac{2S}{L}\right)\right).
\end{multline}

Next, according to the saddle point approximation, we have to sum
over the contributions of all saddle points i.e. to sum over
$\omega$,
\begin{equation}\label{sum}
  \e^{S_{\rm eff}}=\sum_{\omega =1}^{\infty}\e^{S_{\omega}}.
\end{equation}

Some comments are in order. The sum \eqref{sum} diverges, but let us
recall that the validity of our approach was given by the
approximation of smooth eigenvalue distribution. This is definitely
not true for the frequencies $\omega\gtrsim N$, since they involve
wavelengths shorter than the average distance between the
eigenvalues $\theta_n$. So, the applicability of the approach is
restricted to still relatively large values of $\mu$ and $x$. Under
relatively large we understand much bigger than $\sim 1/N$. Also the
equation \eqref{eom-alpha-exp} implies \eqref{eom-mode} only in the
limit $\mu N\gg 1$. Indeed, for finite $N$ the sines/cosines of
$\omega\gtrsim N$ can be expressed in terms of sines/cosines of
smaller arguments. Therefore, \eqref{eom-mode} will get corrections
from the higher modes: $\omega\gtrsim N$. For $\mu\gg 1/N$ this
contribution is exponentially suppressed and we can neglect the
modes higher than $N$, otherwise they should be taken into account.
As a conclusion the sum in \eqref{sum} should be understood as
regularized by restricting it to $\omega\leq N$.

It is clear, however, that as soon as large numbers are concerned,
$S_{\omega}\gg 1$ and we have,
\begin{equation}
  \e^{S_{1}}\gg\e^{S_{\omega}}=[\e^{S_{1}}]^{\ft1\omega},\qquad \forall \omega
  >1.
\end{equation}
Hence, for large $L$ all terms except the first one can be neglected
and we have $S_{\rm eff}=S_1$.

As one can see from \eqref{noN}, there is no $N$ and $N^2$ scaling
left in the effective action. It would be also interesting to check
the effective action \eqref{noN} against the PET result.

% ========================================
\subsubsection{Solution for small $\mu$ and $x$}\label{sec:small-mu}
Small chemical potentials correspond to larger expectation values
for respective thermodynamical quantities, since the suppression of
configurations with large charges is weaker. Therefore considering
small values of $\mu$ and $x$ will favor contributions with large
$L$ and, respectively, $S$.

In the approach of previous subsection, however, this limit is
singular since the expansion in powers of $\e^{-\mu_{\pm}}$ diverges
for small $\mu_{\pm}$.

Now, since $\mu_{\pm}$ are small we expect that the eigenvalues
$\theta_n$ should condense around some arbitrary point breaking the
translational symmetry. Let us pass to the quantitative analysis of
this case as well.

For small values of $\mu_{\pm}$ the trigonometric and hyperbolic
cosines can be replaced by the first terms of their Taylor
expansion, $1\pm x^2/2+\dots$. In this limit the action assumes the
following form,
\begin{multline}\label{s-alpha}
  S(\theta;\mu,\vec{x})=-N^2(\mu-\ln 2)+N[\ln(\mu_{+}\mu_{-})-2\ln 2]\\
  +\ft12\sum_{\substack{m,n\\
  n\neq  m}}\left(-\ln\theta^2_{mn}+\ln(\mu_{+}^2+\theta^2_{mn})
  +\ln(\mu^2_{-}+\theta^2_{mn})\right).
\end{multline}

The ``action'' \eqref{s-alpha} leads to the following ``equations of
motion'',
\begin{equation}\label{eq-discr}
   \sum_{m\neq
   n}\left(-\frac{1}{\theta_{nm}}+
   \frac{\theta_{nm}}{\mu_{+}^2+\theta_{nm}^2}+
   \frac{\theta_{nm}}{\mu_{-}^2+\theta_{nm}^2}\right)=0.
\end{equation}
Let us note, that when at least one of $\mu_{\pm}$ vanishes, the
potential becomes purely attractive and all eigenvalues collapse to
a single point. In general, potential becomes attractive outside the
region $\theta_{mn}\lesssim \sqrt{\mu_{+}\mu_{-}}$, so the
eigenvalues should condense to a region of this typical size.

The problem of finding the eigenvalue configuration satisfying
\eqref{eq-discr} is equivalent to finding an equilibrium
distribution of particles on a line with a Van-der-Waals-like pair
interaction given by the potential
\begin{equation}\label{potential}
  \varphi(\lambda)=\ft12\left(-\ln\lambda^2+
  \ln(\mu_{+}^2+\lambda^2)+\ln(\mu_{-}^2+\lambda^2)\right).
\end{equation}

Again, in the limit of large $N$, the the eigenvalues become dense
and one can replace the equation \eqref{eq-discr} by a continuous
integral one:
\begin{equation}\label{eq-int}
  \int_{-\infty}^{\infty}\dd \eta\, \rho(\eta)
  \left(-\frac{1}{\lambda-\eta}+
  \frac{\lambda-\eta}{\mu_{+}^2+(\lambda-\eta)^2}+
  \frac{\lambda-\eta}{\mu_{-}^2+(\lambda-\eta)^2}
  \right)=0,
\end{equation}
where the equation should hold for $\lambda\in\supp \rho(\lambda)$.

We approach this problem in the Appendix \ref{app:elst}. The
resulting `energy' function \eqref{final-e} generates following
Legendre equations\footnote{Note that here and on we replaced the
``original'' length $L$ by a ``renormalized'' one $L\to L-N^2$.}:
\begin{subequations}\label{leg-e}
\begin{align}\label{leg-el}
  L&=2N^2\left(\mu_{+}^{-1}-\mu_{-}^{-1}\right)
  \left.\xi \frac{\pd\E}{\pd \xi}\right|_{\xi=\sqrt{\frac{\mu_{+}}{\mu_{-}}}},\\
  \label{leg-es}
  2S&=2N^2\left(\mu_{+}^{-1}+\mu_{-}^{-1}\right)\left.\xi \frac{\pd\E}{\pd
  \xi}\right|_{\xi=\sqrt{\frac{\mu_{+}}{\mu_{-}}}}
  .
\end{align}
\end{subequations}
where the function $\E$ is given by (see \eqref{final-e} in the
Appendix \ref{app:elst}),
\begin{equation}
  \E(\xi)=\xi\arctan\xi^{-1}+\xi^{-1}\arctan\xi,\qquad
  \xi=\sqrt{\frac{\mu_{+}}{\mu_{-}}}.
\end{equation}

In terms of $L_{\pm}=L\pm 2S$ Legendre equations \eqref{leg-el} take
the form,
\begin{equation}\label{l-pm}
  L_{\pm}=\pm 4N^2\mu_{\pm}^{-1}\xi\E'(\xi).
\end{equation}

Dividing the equation for ``$+$" by the one for ``$-$", we obtain,
\begin{equation}\label{mu-pm-sol}
  \xi^2\equiv\frac{\mu_{+}}{\mu_{-}}=-\frac{L_{-}}{L_{+}}.
\end{equation}

A remark is in order. Physically meaningful region is one where
$0\leq L_{-}\leq L\leq L_{+}$. This corresponds to such a regime
where chemical potentials statistically suppress the spin stronger
then the length $L$. On the other hand, we know that the matrix
integral diverges at $\mu\leq x/2$. So, to reach the physically
relevant region we are forced to make an analytic continuation to
complex $x$ and $\xi^2<0$.

Eq. \eqref{l-pm} together with \eqref{mu-pm-sol} yield,
\begin{equation}
  \mu_{\pm}=\pm 4N^2L_{\pm}^{-1}\xi\E'(\xi),
\end{equation}
while in terms of $\mu$ and $x$ this solution takes the form
\begin{align}\label{sol-mu}
  \mu&=2N^2(L_{+}^{-1}-L_{-}^{-1})\xi \E'(\xi),\\
  \label{sol-xi}
  x&=2N^2(L_{+}^{-1}-L_{-}^{-1})\xi \E'(\xi)
\end{align}

Plugging this solution into the Legendre transform equation
\eqref{legendre}, we obtain,
\begin{multline}\label{s-eff-fin}
  S_{\rm eff}(L,S)\equiv\ft12(L_{+}\mu_{+}+L_{-}\mu_{-})-4N^2\E(\xi)=
  -4N^2\E(\xi)=\\
  4 N^2
  \left(\sqrt{\frac{L_+}{L_-}} \ln\left|\frac{1+
  \sqrt{\frac{L_-}{L_+}}}{1-\sqrt{\frac{L_-}{L_+}}}\right|
  +\sqrt{\frac{L_-}{L_+}}\ln\left|\frac{1+
  \sqrt{\frac{L_+}{L_-}}}{1-\sqrt{\frac{L_+}{L_-}}}\right|
  \right),
\end{multline}
where $L_{\pm}$ are given by,
\begin{equation}
  L_{\pm}=L\pm 2S.
\end{equation}

Let us recall that $\e^{S_{\rm eff}(L,S)}$ gives the number of composite SYM operators of length $L$ and spin $S$. It is instructive to compare eq. \eqref{s-eff-fin} to the log of the number of states of given spin in a system of $L$ completely random spins.

% =====================================
\newpage
\subsection{One-loop contribution}
So far we considered the purely oscillator part of the matrix model.
As we discussed above in the limit of high temperature and small
coupling the quartic commutator term can be considered to be a small
perturbation. Therefore, in this limit one can find the correction
to the oscillator effective action by just evaluating the average of
the quartic term in the oscillator dominated background. This
amounts to neglecting among others the back reaction of the
background to the presence of the perturbation. At the one loop
level it is a justified approximation, since the back reaction comes
at the order $g_{\rm YM}^4$ and are of the same order as
\emph{two-loop} contribution from SYM.\footnote{Taking into account
corrections from higher orders could be similar to one-loop
renormalization group improvement.}

In the previous section we considered the quadratic part
\eqref{ham0} of the matrix model Hamiltonian \eqref{ham},
\begin{multline}
  H\equiv H_0+V=\\
  \tr\left(\mu\bar{\Psi}_a \Psi_a+\ft12
  \vec{x}\cdot\bar{\Psi}_a\vec{\sigma}_{ab} \Psi_b\right)
  -\frac{\beta g_{\rm YM}^2}{16\pi^2}\tr[\bar{\Psi}_a,
  \bar{\Psi}_b][\Psi^a,\Psi^b].
\end{multline}

At high temperatures the coupling $\beta g_{\rm YM}^2$ is small so
we can expand the partition function as follows
\begin{multline}\label{z-corr}
  Z=\e^{-S_{\rm eff}}=\tr \e^{-(H_0+V)}\\
  =\tr\e^{-H_0}-\tr\e^{-H_0}V+\dots=Z_0(1-\langle V \rangle_0+\dots)\\
  \approx\e^{-S_{\rm eff}^{(0)}-\langle V \rangle_0} ,
\end{multline}
where the mean value $\langle V \rangle_{0}$ is defined in a
standard way
\begin{equation}\label{v:ev}
  \langle V \rangle_0=Z_0^{-1}\tr\e^{-H_0}V.
\end{equation}
Thus, computation of non-zero coupling correction at the leading
order resides in the calculation of the mean \eqref{v:ev} of the
following operator
\begin{equation}\label{V}
  V=\frac{\beta g_{\rm YM}^2}{16\pi^2}\tr[\bar{\Psi}_a,
  \bar{\Psi}_b][\Psi^a,\Psi^b].
\end{equation}
To do it let us use again the anti-holomorphic representation (see
Appendix \ref{app:holomorphic}).

Since the operator \eqref{V} is a normal one its anti-holomorphic
kernel is given by Eq. \eqref{kernel-norm},
\begin{equation}
  K_V(\bar{\Phi},\Phi)=\frac{\beta g_{\rm YM}^2}{16\pi^2}
  \e^{\tr\bar{\Phi}_a\Phi_a}\tr[\bar{\Phi}_a,\bar{\Phi}_b][\Phi_a,\Phi_b].
\end{equation}

Since both the bare quadratic Hamiltonian and the perturbation are
gauge invariant it suffices to insert the projection to the gauge
invariant subspace only once. Therefore, the trace \eqref{v:ev} is
given by
\begin{multline}\label{gi-tr-1loop}
  Z_0\langle V\rangle_0=\\
  \frac{\beta g_{\rm YM}^2}{8\pi^2}
  \int\dd \theta\,\Delta(\theta)
  \int\frac{\dd\bar{\Phi}\dd\Phi}{\pi^{N^2}}\,
  \exp\left(-\sum_{mn}\left(h^{(+)}_{mn}\bar{\Phi}^{(+)}_{mn}\Phi^{(+)}_{mn}
  +h^{(-)}_{mn}\bar{\Phi}^{(-)}_{mn}\Phi^{(-)}_{mn}\right)\right)\times \\
  \tr[\bar{\Phi}_{+}\e^{-\mu_{+}},\bar{\Phi}_{-}\e^{-\mu_{-}}][\Phi_{+},\Phi_{-}],
\end{multline}
where $h^{(\pm)}_{mn}$ was defined in \eqref{h-pm} of subsection
\ref{sec:quadratic}.

Making the substitution:
$h^{(\pm)}_{mn}\bar{\Phi}^{(\pm)}_{mn}\to\Phi^{(\pm)}_{mn}$ the
integral \eqref{gi-tr-1loop} transforms into
\begin{equation}\label{V:ah}
  \langle V\rangle_0=
  \frac{\beta g_{\rm YM}^2}{8\pi^2}
  \int\frac{\dd\bar{\Phi}\dd\Phi}{\pi^{N^2}}\, \e^{-\bar{\Phi}\Phi}
  \e^{-2\mu}\tr[(\bar{\Phi}_{+}/h^{(+)}),
  (\bar{\Phi}_{-}/h^{(-)})]
  [\Phi_{+},\Phi_{-}],
\end{equation}
where $(\bar{\Phi}_{\pm}/h^{(\pm)})$ is a matrix given by the
elements
\begin{equation}
  (\bar{\Phi}_{\pm}/h^{(\pm)})_{mn}=\bar{\Phi}^{\pm}_{mn}/h^{(\pm)}_{mn},
\end{equation}
and we used $Z_0=1/\prod h^{(+)}h^{(-)}$ to cancel $Z_0$ in the
l.h.s.

This matrix integral can be computed taking into account that,
\begin{multline}\label{int:I}
  I_{k_1l_1}^{a_1}{}_{k_2l_2}^{a_2}{}_{m_1n_1}^{b_1}{}_{m_2n_2}^{b_2}\equiv
  \int\frac{\dd\bar{\Phi}\dd\Phi}{\pi^{N^2}}\,
  (\bar{\Phi}_{k_1l_1}^{a_1}\bar{\Phi}_{k_2l_2}^{a_2}
  \Phi_{m_1n_1}^{b_1}\Phi_{m_2n_2}^{b_2})\,\e^{-\tr\bar{\Phi}\Phi}=\\
  \delta^{a_1 b_1}\delta^{a_2 b_2}\delta_{n_1 k_1}\delta_{m_1 l_1}
  \delta_{n_2 k_2}\delta_{m_2 l_2}+
  \delta^{a_1 b_2}\delta^{a_2 b_1}\delta_{n_1 k_2}\delta_{m_1 l_2}
  \delta_{n_2 k_1}\delta_{m_2 l_1}.
\end{multline}

Now, plugging \eqref{int:I} into \eqref{V:ah} we get,
\begin{multline}\label{V:h}
  \langle V\rangle_0=\frac{\beta g_{\rm YM}^2}{4\pi^2}\e^{-2\mu}
  \left(\sum_{n}\frac{1}{h^{(+)}_{nn}h^{(-)}_{nn}}
  -\ft12\sum_{knm}\left(\frac{1}{h^{(+)}_{mn}h^{(-)}_{nk}}
  +\frac{1}{h^{(-)}_{mn}h^{(+)}_{nk}}\right)\right)\\
  =\frac{\beta g_{\rm YM}^2}{4\pi^2}
  \left(\frac{N}{(\e^{\mu_{+}}-1)(\e^{\mu_{-}}-1)}\right.\\
  -\left.\ft14\sum_{knm}\frac{\cos\theta_{mk}-
  \e^{-\mu_{+}}\cos\theta_{nk}-\e^{-\mu_{-}}\cos\theta_{mn}+\e^{-2\mu}}
  {(\cosh\mu_{+}-\cos\theta_{mn})(\cosh\mu_{-}-\cos\theta_{nk})}\right)\\
  \equiv\widetilde{\langle V\rangle_0}+\langle V\rangle_0'.
\end{multline}

The term in the first line of the r.h.s. of \eqref{V:ah} can be
readily evaluated to be,
\begin{equation}\label{tildeV}
  \widetilde{\langle V\rangle_{0}}(\mu,\vec{x})=\frac{\beta g_{\rm YM}^2N}{4\pi^2}
  \frac{\e^{-\mu_{+}-\mu_{-}}}{(1-\e^{-\mu_{+}})(1-\e^{-\mu_{-}})},
\end{equation}
or in terms of the level $\omega$ solution to the Legendre equations
(see \eqref{leg-sol}):
\begin{equation}
  \mu_{\pm}=\frac{1}{\omega}\ln\left(\frac{2L}{L\mp 2S}\right),
\end{equation}
the same quantity can be written as,
\begin{equation}\label{wave-mean}
  \widetilde{\langle V\rangle_{0}}(L,\vec{S})=
  \frac{\beta \lambda}{4\pi^2}
  \frac{1}{\left(\left(\frac{2L}{L-2S}\right)^{\ft1\omega}-1\right)
  \left(\left(\frac{2L}{L+2S}\right)^{\ft1\omega}-1\right)}.
\end{equation}
The $\omega=1$ term is a constant,
\begin{equation}\label{wave-mean-1st}
  \widetilde{\langle V\rangle_{0}}=
  \frac{\beta \lambda}{4\pi^2}.
\end{equation}

It is interesting to note that for large $\omega$ the mean value
\eqref{wave-mean} behaves like,
\begin{multline}
  \widetilde{\langle V\rangle_{0}}=\\
  \frac{\beta
  \lambda}{4\pi^2}\left\{
  \frac{\omega^2}{\ln\left(\frac{2L}{L-2S}\right)\ln\left(\frac{2L}{L+2S}\right)}
  -\frac{\omega}{2}\left(\frac{1}{\ln\left(\frac{2L}{L+2S}\right)}+
  \frac{1}{\ln\left(\frac{2L}{L+2S}\right)}\right)
  +\dots\right\}.
\end{multline}
The overall negative sign in front of $\omega^2$ term makes the
terms with large $\omega$ to be even more strongly suppressed.

\subsubsection{Expansion in powers of $\e^{-\mu_{\pm}}$}
Let us consider the second line of  \eqref{V:ah} and expand it in
powers of $\e^{-\mu_{\pm}}$ the same way as we did in the previous
section. Remarkably, again the expansion is not only doable, but
also appears in a simple form,
\begin{multline}\label{v-mean-anlt}
  \langle V\rangle_0'=\langle V\rangle_0-\widetilde{\langle
  V\rangle_{0}}=\\
  -\frac{\beta g_{\rm YM}^2}{4\pi^2}
  \sum_{\omega,\omega'}
  \e^{-(\omega+1) \mu_{+}-(\omega'+1) \mu_{-}}
  \sum_{mnk}\cos(\omega\theta_{mn}+\omega' \theta_{kn})
  ,
\end{multline}
where $\quad \omega,\omega'=1,2,\dots$ enumerate the modes of the
eigenvalue density distribution.

As a warmup it is not difficult to see that for homogeneous
distribution of eigenvalues $\theta_n$ the average of each such term
vanishes i.e.,
\begin{equation}\label{lg-V-fin-large-mu}
  \langle V\rangle_0=\widetilde{\langle V\rangle_{0}}.
\end{equation}

Now let us turn to the situation where the eigenvalue distribution
is `perturbed'  by a non-constant mode $\omega$.

A straightforward way to compute the expectation value $\langle
V\rangle_0$ would be to plug the eigenvalue distribution into
\eqref{v-mean-anlt} and get the answer. In the case of analytic
solution we have the possibility of an indirect way of evaluation of
\eqref{v-mean-anlt} even without knowledge of the details relating
to the density $\rho(\lambda)$.

Indeed, let us expand the cosine in \eqref{v-mean-anlt} and use the
even character of the eigenvalue distribution (see the Appendix
\ref{app:cos}) to get,
\begin{multline}\label{V:modes}
  \langle V\rangle_0'=-\frac{\beta g_{\rm YM}^2}{4\pi^2}\e^{-2\mu}
  \sum_{\omega,\omega'}\e^{-\omega\mu_{+}-\omega\mu_{-}}
  \tilde{\rho}_{\omega}\tilde{\rho}_{\omega'}\tilde{\rho}_{\omega+\omega'}=
  -\frac{\beta g_{\rm YM}^2}{4\pi^2}\e^{-2\mu}\times\\
  \left(
  \sum_{\omega}\left(\e^{-\omega\mu_{+}}+\e^{-\omega\mu_{-}}\right)
  \rho_0\tilde{\rho}_{\omega}^2+
  \sum_{\substack{\omega,\omega'\\
  \neq  0}}\e^{-\omega\mu_{+}-\omega\mu_{-}}
  \tilde{\rho}_{\omega}\tilde{\rho}_{\omega'}\tilde{\rho}_{\omega+\omega'}
  \right),
\end{multline}
where we separated explicitly the zero mode $\rho_0=N/2\pi$ of the
eigenvalue distribution density. Using the property \eqref{irr} of
the modes of the density one can readily evaluate the first term of
the last equality of \eqref{V:modes} to be,
\begin{multline}\label{V:first}
  \langle V\rangle_0'=-\frac{\beta g_{\rm YM}^2}{4\pi^2}\e^{-2\mu}
  \sum_{\omega}\left(\e^{-\omega\mu_{+}}+\e^{-\omega\mu_{-}}\right)
  \rho_0\tilde{\rho}_{\omega}^2=\\-\frac{\beta g_{\rm YM}^2N}{8\pi^3}\e^{-2\mu}
  \sum_{\omega}\tilde{\rho}_{\omega}^2
  =-\frac{\beta \lambda}{(2\pi)^3}\e^{-2\mu}L=\\
  -\frac{\beta
  \lambda}{(2\pi)^3}L\left(1-\left(\frac{2S}{L}\right)^2\right)^{\frac{1}{\omega}},
\end{multline}
or for $\omega=1$ it becomes,
\begin{equation}
  \langle V\rangle_{0,\omega=1}'=-\frac{\beta
  \lambda}{(2\pi)^3}\left(L^2-4S^2\right)/L.
\end{equation}
By contrast, for $\omega\to\infty$, this is,
\begin{equation}
  \langle V\rangle_{0,\omega\to\infty}'=-\frac{\beta
  \lambda}{(2\pi)^3}L.
\end{equation}

As about the second term of \eqref{V:modes}, it should vanish, since
there is only one non-trivial mode $\omega$.

\subsubsection{Solution for small $\mu$ and $x$}

Consider now the case of small $\mu$ and $x$. In this case the only
available approach is to plug the solution for the eigenvalue
density \eqref{density} into \eqref{V:h} to evaluate $\langle
V\rangle_{0}$.

The term $\widetilde{\langle V\rangle_{0}}$ in this approach is of the order $N^0$, and, therefore is small with respect to the terms $\langle V\rangle_{0}'$.

Let us turn to the evaluation of $\langle V\rangle_{0}'$. The
leading term in $\mu_{\pm}$ and $\theta_{mn}$ of the second line of
\eqref{V:h} reads,
\begin{multline}\label{V:elst}
  \langle V\rangle_{0}'=\frac{\beta g_{\rm YM}^2}{4\pi^2}
  \sum_{mnk}\frac{-\theta_{mn}\theta_{nk}+\mu_{+}\mu_{-}}
  {(\mu_{+}^2+\theta_{mn}^2)(\mu_{-}^2+\theta_{nk}^2)}=\\
  \frac{\beta g_{\rm YM}^2}{4\pi^2}
  \int\dd\lambda\,\rho(\lambda)\int\dd\eta\,\rho(\eta)\int\dd\nu\,\rho(\nu)\\
  \times
  \frac{\mu_{+}\mu_{-}-(\eta-\lambda)(\lambda-\nu)}{(\mu_{+}^2+(\eta-\lambda)^2)
  ((\mu_{-}^2+(\nu-\lambda)^2)}
  .
\end{multline}

In the incompressible liquid approximation of the
Appendix~\ref{app:elst}, the integral in \eqref{V:elst} becomes,
\begin{multline}\label{loglog}
  \langle V\rangle_{0}'=N^2\frac{\beta \lambda}
  {2(2\pi)^2\Lambda}\int_{-\Lambda}^{\Lambda}\dd\lambda\,\times\\
  \left\{
  -\ln\left(\frac{\lambda^2-(\Lambda+\ii\mu_{+})^2}
  {\lambda^2-(\Lambda-\ii\mu_{+})^2}\right)
  \ln\left(\frac{\lambda^2-(\Lambda+\ii\mu_{-})^2}
  {\lambda^2-(\Lambda-\ii\mu_{-})^2}\right)\right.\\
  \left.+\ft14\ln\left(\frac{\mu_{+}^2+(\lambda+\Lambda)^2}
  {\mu_{+}^2+(\lambda-\Lambda)^2}\right)
  \ln\left(\frac{\mu_{-}^2+(\lambda+\Lambda)^2}
  {\mu_{-}^2+(\lambda-\Lambda)^2}\right)\right\},
\end{multline}
where we use the shortcut notation $\Lambda=\sqrt{\mu_{+}\mu_{-}}$.
Performing a substitution $\lambda\to\lambda\Lambda$ in
\eqref{loglog} we get for the correction,
\begin{equation}
  \langle V\rangle_{0}'=N^2\frac{\beta \lambda}
  {2(2\pi)^2}F\left(\sqrt{\frac{L_{-}}{L_{+}}}\right),
\end{equation}
where the function $F(\xi)$ is given by the integral,
\begin{multline}\label{F}
  F(\xi)=\int_{-1}^{1}\dd\lambda\,
  \left\{
  -\ln\left|\frac{\lambda^2-(1-\xi)^2}{\lambda^2-(1+\xi)^2}\right|
  \ln\left|\frac{\lambda^2-(1+\xi^{-1})^2}
  {\lambda^2-(1-\xi^{-1})^2}\right|\right.\\
  +\left.\ft14
  \ln\left|\frac{-\xi^2+(\lambda+1)^2}{-\xi^2+(\lambda-1)^2}\right|
  \ln\left|\frac{-\xi^{-2}+(\lambda+1)^2}{-\xi^{-2}+(\lambda-1)^2}\right|
  \right\},
\end{multline}
which can be computed explicitly and is a combination (rather long)
of log and poly-log functions. The integration over the main branch
of the log is assumed. $F(\xi)$ has the following properties: It is
invariant with respect to inversion: $\xi\to\xi^{-1}$. Also
$F(0)=F(\infty)=0$ and it has a maximum at one.

\newpage
% ---------------------------------------------------------
\section{Spin chain gas at low temperature}\label{sec:lowT}
Consider the Hamiltonian \eqref{ham} in the regime when the inverse
temperature $\beta$ is large. In this regime the commutator term in
the Hamiltonian becomes dominating over the quadratic part. If at
the same time the Yang--Mills coupling is small enough, the one loop
term becomes dominant also over the higher loop contributions. As
discussed in \cite{Sochichiu:2005jp} (see \cite{Sochichiu:2005ex}
for the background), at the large value of coupling in front of the
commutator term causes configurations with non-vanishing commutator
to be statistically suppressed. In the extremal case, when $\beta
g_{\rm YM}^2\to\infty$, the field is forced to remain in the valley of
the potential i.e. the allowed space will be restricted to matrices
satisfying,
\begin{equation}
  \tr[\bar{\Psi}_a,\bar{\Psi}_b][\Psi_a,\Psi_b]=0.
\end{equation}
This condition is equivalent to restriction to commuting matrices,
\begin{equation}\label{psi-comm}
  [\Psi_a,\Psi_b]=0.
\end{equation}

Since $\Psi_a$ are not Hermitian, the condition \eqref{psi-comm} is not enough for them to be diagonalizable. The Gauss law constraint,
\begin{equation}
  G\equiv[\bar{\Psi}_a,\Psi_a]=0,\qquad a=1,2
\end{equation}
implies, however, that the commuting matrices are also normal. Even more generally, they should satisfy, \begin{equation}\label{norm-comm}
 [\bar{\Psi}_a,\Psi_{b}]=0.
\end{equation}
The condition \eqref{norm-comm} is already sufficient for simultaneous diagonalization of all $\Psi_a$ as well as $\bar{\Psi}_a$ whose eigenvalues are complex conjugate to those of $\Psi_a$.

Passing to description in terms of eigenvalues $\Psi^a_n$ of
$\Psi_a$ as well as $\bar{\Psi}^a_n$ of $\bar{\Psi}_a$ in quantum
theory can be associated with change of variables in the path
integral. The Hamiltonian takes the form,
\begin{equation}
  H=\mu\tr\bar{\Psi}_a\Psi_a+\ft12
  \vec{x}\cdot\vec{\sigma}_{ab}\tr\bar{\Psi}_a\Psi_b=
  \mu_{+}\tr\bar{\Psi}_{+}\Psi_{+}
  +\mu_{-}\tr\bar{\Psi}_{-}\Psi_{-},
\end{equation}
where in the last equality we used the eigenbasis of matrix
$(\vec{x}\cdot \vec{\sigma})$. The change of variables to diagonal
values and angular component gives rise to a Jacobian which is the
product of the Vandermonde determinants for all $\Psi$ and
$\bar{\Psi}$,
\begin{equation}
  \Delta^2(\bar{\Psi},\Psi)=\prod_a\prod_{m>n}|\Psi^a_m-\Psi^a_n|^2.
\end{equation}

A common trick  in the Schroedinger equation approach is to absorb
this measure (see e.g. \cite{Brezin:1977sv}) by a redefinition of
the wave function,\footnote{Note, however, that in this case the
redefinition is an operator action.}
\begin{equation}
  \psi\mapsto\sqrt{\Delta^2(\bar{\Psi},\Psi)}\psi.
\end{equation}
The continuity of the wave function requires that the sign of the root is chosen in such a way that after the redefinition the wave function becomes antisymmetric in diagonal modes. From the point of view of path integral approach vanishing of the measure at such points means that the configurations with coinciding eigenvalues are prohibited since they have zero probability density.

As large $N$ or large $L$ are concerned the number of excluded
states is negligible with respect to the total number of states.
Then, the partition function can be evaluated by the direct
computation of the trace,
\begin{multline}\label{z-diag}
  Z(\mu,\vec{x})=\tr\e^{-H(\mu,\vec{x})}=
  \sum_{\{N_n^{\pm}=0,1,2\dots\}}
  \e^{-\mu_{+}\sum_n N_n^{+}-\mu_{-}\sum_n N_n^{-}}=\\
  \left[ \frac{1}{1-\e^{-\mu_{+}}}\frac{1}{1-\e^{-\mu_{-}}}\right]^N
  =\exp\left( -N[\ln(1-\e^{-\mu_{+}})-\ln(1-\e^{-\mu_{-}})]\right).
\end{multline}
As we discussed earlier, large $L$ and $\vec{S}$ correspond to small values of the chemical potentials $\mu_{\pm}$. In this limit the partition function becomes,
\begin{equation}
  Z(\mu,\vec{x})=\exp\left( -N\ln\mu_{+}\mu_{-}\right).
\end{equation}

Solution to the Legendre transform equations \eqref{leg} gives the
following expression for the effective action,
\begin{equation}
  S_{\rm eff}(L,\vec{S})=N\ln L_{+}L_{-}=N\ln (L^2-4\vec{S}^2).
\end{equation}

% ---------------------------------------------------------
\section{String theory interpretation}

Let us give a qualitative string theory interpretation of above results. We considered the effective action which is the thermodynamical potential counting the number of SYM operators of a given length $L$ and given spin $S$. According to the AdS/CFT dictionary this corresponds to the entropy of AdS (multi)string states having $S$ unites of angular momentum in the (12) plane and $L-S$ unites of the angular momentum in the (56) plane of $S^5$, regarded as embedded into $\mathbb{R}^6$.
(This corresponds to the choice $\phi\propto (\phi_1+\ii\phi_2)$ and $Z\propto (\phi_5+\ii \phi_6)$.)

Basically, we found three different regimes in the behavior of the model which are distinguished by different scaling of the thermodynamical quantities at large $N$: I. $N$-independent, II. linear $N$-scaling and, finally, III. quadratic $N$-scaling. Recall that the meaning of $N$ on the string side is the number of $D3$-branes providing us with AdS$_5\times S^5$ background in their vicinity. Consider all three cases in detail.
\begin{itemize}
  \item[I.] The effective number of degrees of freedom does not depend on $N$ rather on $L$ and $S$. This means that we are in a genuine stringy phase as strings with no internal structure present the effective fundamental excitations which we are counting. This phase occurs at moderate values of $L$: $L\lesssim \sqrt{N}$.
  \item[II.] Linear $N$-dependence says that we have a number of degrees of freedom proportional to the number of $D3$-branes. Therefore, it is natural to assume, that it is the $D3$-branes which are the fundamental excitations of the theory in this phase. This phase corresponds to large values of $\beta g_{\rm YM}^2$ and moderate $L$. (To have the one-loop contribution leading one has to require also that $\beta g_{\rm}^4$ is small.)
  \item[III.] The $N^2$ scaling, generally means that the $D3$-brane is not anymore a fundamental excitation but can be considered as a condensate or bound state of smaller objects. The situation is very similar to brane condensation in noncommutative gauge theories \cite{Sochichiu:2000ud,Sochichiu:2000bg,Sochichiu:2000kz,Kiritsis:2002py} (for a review see also \cite{Sochichiu:2002jh}). This occurs at large $L$: $L>\sqrt{N}$. Remarkably, the model in this regime is best described in terms of noncommutative gauge theory \cite{Bellucci:2004fh}.
\end{itemize}
% ---------------------------------------------------------
\section{Discussion}

In this paper we considered a non-Hermitian matrix model describing the anomalous dimension spectrum of $\mathcal{N}=4$ SYM theory. In this study we did not recur to the integrability at infinite $N$ neither to the description in terms of spin system. Moreover, $N$ all the time was kept a finite statistically large number. We introduce a notion of temperature, which in the SYM theory plays the role of auxiliary parameter allowing restoration the density of the anomalous dimensions of composite SYM operators. At the same time due to the fact that the dilatation operator is the Hamiltonian corresponding to the radial time in conformal theory this temperature corresponds to compactification of the SYM to the four-dimensional sphere $S^4$.  In the framework of string/gauge duality this temperature is the genuine temperature of the dual string theory and can be attributed to the presence of a black hole \cite{Witten:1998zw}.

For the two-matrix model we succeeded to compute the thermal partition function and the effective action depending on the total occupation number, which in the spin language is the aggregate spin chain length, and the occupation number difference, which corresponds to the total spin. We did this in the both approximation of high as well as of low temperature.
 The most attention and effort was given to the high temperature regime. In this regime we found a rather elegant approach based on the analytic expansion in terms of exponentials of chemical potentials. This approach allows an \emph{exact} saddle point evaluation of the partition function. For sufficiently large chemical potentials $\mu_{\pm}$ the contribution to thermodynamical potentials which scale like $N^2$ and $N$ does not appear. This is compatible with the PET (P\'olya Enumeration Theorem) evaluations which holds for $\mu$ above the Hagedorn critical value. As in our case the spin is also included we may conjecture that the phase separation should occur along the critical line given by\footnote{This is also supported by the random walk approach \cite{Sochichiu:2006yv}.}: \[ 1-\e^{-\mu_{+}}-\e^{-\mu_{-}}=0, \] at which the first non-trivial mode to the saddle point equation appear. (In the case of zero $x$ the above critical line reduces exactly to the Hagedorn criticality condition for $\mu$, note that $\mu$ for the oscillator has the same meaning as the temperature.) In this regime the fundamental excitations are stringy-like.

Using the incompressible liquid model approximation for the gauge field eigenvalue condensate we found a description of the model at small chemical potentials $\mu_{\pm}$, which can be trusted at least qualitatively. For this regime we found a $N^2$ behavior, which corresponds to string bit phase. This phase is characterized by melting of strings into point objects: \emph{string bits}. As a result the model looks as a system of $N^2$ interacting particles.

At small temperatures the matrix fluctuations are bind to the diagonal. Therefore the effective number of degrees of freedom is proportional to $N$. As $N$ is the number of $D3$-branes in the string theory, the natural assumption is that they are the elementary excitations in this regime.

% ---------------------------------------------------------
\subsection*{Acknowledgements}
I benefited from useful discussions with many colleagues, in
particularly it is my pleasure to acknowledge ones with J.
Erdmenger, R. Ferrari, J. Gro{\ss}e, J.-H. Park, E. Seiler, N.
Slavnov and P. Weisz.

This work was possible to complete due to the extension of my A. von Humboldt fellowship program in Max-Planck-Institute f\"{u}r Physik in Munich. I am grateful to the staff and visitors of MPI for the friendly and creative atmosphere I profited during my stay in Munich. Special thanks are for my host, Dieter Luest, for interest and support.

\newpage
\appendix
\section{Anti-holomorphic representation}\label{app:holomorphic}

A very useful representation of wave functions of quantum
oscillators is given by the anti-holomorphic representation. For the
convenience of the Reader and to fix the notations we give here a
brief account of this representation.

Consider the oscillator algebra generated by the ladder operators
$a$ and $a^\dag$ which are subject to the commutation relations
\begin{equation}\label{ho}
  [a,a^\dag]=1.
\end{equation}

We can represent the quantum states in the theory as
anti-holomorphic function $f(\bar{z})$ of a complex variable $z$.
The action of the ladder operators on such states is given by
\begin{equation}
  a^\dag\to \bar{z},\qquad a\to \pd/\pd\bar{z},
\end{equation}
i.e. the raising operator $a^\dag$ acts by multiplication by
$\bar{z}$, while the lowering $a$ acts as a derivative. Obviously
the commutation relation \eqref{ho} is satisfied by such definition.

Now it is not very difficult to recoverer all formulas used to
construct the representation. The oscillator vacuum is given by the
zero mode of the lowering operator
\begin{equation}
  a\ket{0}=0=\pd \varphi_0/\pd\bar{z}\Rightarrow
  \varphi_0=\constant.
\end{equation}
We can choose $\varphi_0=1$. In this case the orthonormal set of the
oscillator hamiltonian eigenstates is represented by
\begin{equation}
  \ket{n}\sim \frac{\bar{z}^n}{\sqrt{n!}},
\end{equation}
from which we can immediately extract the scalar product rule for
arbitrary (anti-holomorphic) states described by $f(\bar{z})$ and
$g(\bar{z})$:
\begin{equation}\label{sc-pr}
  \bracket{f}{g}=\int\frac{\dd\bar{z}\dd z}{\pi}\,
  \e^{-\bar{z}z}f^*(z)g(\bar{z}).
\end{equation}
A ``well defined'' operator $F$ can be represented by an
anti-holomorphic kernel $F(\bar{z},z)$. The scalar product
\eqref{sc-pr} implies the following rule for the kernel
multiplication:
\begin{equation}
  F\cdot G\sim \int\frac{\dd \bar{w}\dd
  w}{\pi}\,\e^{-\bar{w}w}
  F(\bar{z},w)G(\bar{w},z),
\end{equation}
In particular, the trace of operator $F$ is given by
\begin{equation}
  \tr F=\int\frac{\dd\bar{w}\dd w}{\pi}\,\e^{-\bar{w}w}F(\bar{w},w).
\end{equation}

In many cases the normal symbol of an operator is used. Therefore,
it is useful to know how to pass from one description to another.

Consider an operator $F$ represented in the normal form:
\begin{equation}
  F=F_N(a^\dag,a)\equiv \sum_{m,n}f_{m,n}(a^\dag)^m a^n.
\end{equation}
Since the operators $a^\dag$ and $a$ are normal ordered inside
$F_N(a^\dag,a)$ one can treat it as usual function $F_N(\bar{z},z)$
of complex variable $z$. By analyzing the the matrix element of $F$
between two oscillator states $\bra{m}$ and $\ket{n}$ one finds,
that the normal symbol $F_N(a^\dag,a)$ is related to the
anti-holomorphic symbol in the following simple way:
\begin{equation}\label{kernel-norm}
  F(\bar{z},z)=\e^{\bar{z}z}F_N(\bar{z},z).
\end{equation}

The last useful formula we want to give here is the anti-holomorphic
symbol of the exponent of an operator with quadratic normal form:
\begin{equation}
  U=\e^{-\beta H},\qquad H=h a^\dag a,
\end{equation}
e.g. the anti-holomorphic symbol for the evolution operator of the
harmonic oscillator. Again, the direct computation gives,
\begin{equation}\label{U}
  U(\bar{z},z)=\exp\left(\bar{z}\e^{-\beta h} z\right).
\end{equation}

As a simple test let us compute the trace of \eqref{U}. Modulo
vacuum contribution this gives the oscillator partition function,
\begin{equation}\label{part-f0}
  \tr \e^{-\beta H}=\int\frac{\dd \bar{w}\dd w}{\pi}\,
  \exp\left[-\bar{w}(1-\e^{-\beta
  h})w\right]=\frac{1}{1-\e^{-\beta}}.
\end{equation}
The generalization to higher dimensions is straightforward.

\section{Anti-Holomorphic representation with gauge symmetry}
\label{app:gauge-inv}
Let us consider a situation when oscillator
possesses a gauge symmetry,
\begin{equation}
  z\mapsto z^g,\qquad g\in G.
\end{equation}
This pulls back as a $G$-transformation of the Hilbert space,
\begin{equation}
  \psi(\bar{z})\mapsto \psi(\bar{z}^g)\equiv T(g)\psi(\bar{z}).
\end{equation}
where $T(g)$ is the generator of the induced action.
Infinitesimally,
\begin{equation}
 T(g=\e^{u})\approx \I+t(u),\qquad
 t(u)\psi(z)\approx\psi(\bar{z}^{\e^u})-\psi(\bar{z}),
\end{equation}
where $t(u)=t_\alpha(\bar{z}) u^\alpha$ are the generators of
infinitesimal gauge transformations. Generally $t_\alpha$ have the
form,
\begin{equation}
  t_\alpha(\bar{z})=t_\alpha^i(\bar{z})\bar{\pd}_i, \qquad
  \bar{\pd}_i\equiv\frac{\pd}{\pd\bar{z}^i}
\end{equation}
where $t_\alpha^i(\bar{z})$ are just ordinary functions,
\begin{equation}
  t_\alpha^i(\bar{z})=\left.\frac{\pd^2 \bar{z}^{g(u)}}{\pd u^\alpha\pd
  \bar{z}_i}\right|_{g=1}.
\end{equation}

The gauging consists in projecting the Hilbert space to the subspace
of invariant functions, which amounts to finding subspace
\begin{equation}\label{on-shell}
  t_{\alpha}\cdot \psi(\bar{z})=0.
\end{equation}
According to the canonical quantization procedure \eqref{on-shell}
is equivalent to imposing the on-shell constraint condition since
according to the second theorem by E.~Noether, operator $t_\alpha$
correspond to first-class constraint in the classical theory. When
$G$ possesses an invariant Haar measure e.g. is a compact Lie group
solution to \eqref{on-shell} is remarkably simple\footnote{I am
grateful to Jeong-Hyuck Park for pointing my attention to this
possibility.}
\begin{equation}\label{Haar}
  \psi_{\rm gauge~invariant}(\bar{z})=\int_G\dd g\,
  \psi(\bar{z}^g),
\end{equation}
where we integrate over the gauge group using the invariant Haar
measure $\dd g$: $\dd (h^{-1}gh')=\dd g$. Hence the Haar integral
\eqref{Haar} can be used as the projector to gauge invariant
subspace of the Hilbert space. It is not difficult to check that for
an operator with gauge invariant kernel:
$K(\bar{z}^g,z^g)=K(\bar{x},z)$ action commutes with averaging over
the gauge group,
\begin{multline}
  K\cdot\Pi\cdot\psi=\int\frac{\dd\bar{w}\dd
  w}{\pi}\,\e^{-\bar{w}w}K(\bar{z},w)\psi(\bar{w}^g)=\\
  \int\frac{\dd\bar{w}\dd
  w}{\pi}\,\e^{-\bar{w}w}K(\bar{z}^g,w)\psi(\bar{w})=\Pi\cdot
  K\cdot\psi.
\end{multline}
In particular, the gauged analog of \eqref{part-f0} reads,
\begin{multline}
  \tr\Pi\cdot\e^{-\beta H}=
  \int\frac{\dd\bar{w}\dd w}{\pi}\dd
  g\,\exp\left[-\bar{w}w^g+\bar{w}\e^{-h}w\right]\\
  =\int\dd g\,\left[\det\left(T(g)-\e^{-h}\right)\right]^{-1}.
\end{multline}

\section{The trigonometric sums}\label{app:cos}

We use the following trigonometric sums,
\begin{multline}
   \sum_{mnk}\cos\left(\omega \theta_{mn}+\omega' \theta_{kn}\right)=\\
   \sum_{mnk}\cos \left(\omega  \theta_m\right)
   \cos \left(\omega '\theta_k \right)
   \cos \left(\left(\omega +\omega '\right)\theta_n\right)\\
   -\sum_{mnk}
   \sin \left(\omega  \theta_m\right)
   \sin \left(\omega'\theta_k\right)
   \cos \left(\left(\omega+\omega'\right)\theta_n \right)\\
   +\sum_{mnk}
   \sin \left(\omega  \theta_m\right)
   \cos \left(\omega'\theta_k \right)
   \sin \left(\left(\omega +\omega'\right)\theta_n \right)\\
   +\sum_{mnk}
   \cos \left(\omega  \theta _m\right)
   \sin \left(\omega'\theta_k \right)
   \sin \left(\left(\omega +\omega'\right)\theta_n \right)\\
   =\tilde{\rho}_{\omega}\tilde{\rho}_{\omega'}\tilde{\rho}_{\omega+\omega'}
   -\hat{\rho}_{\omega}\hat{\rho}_{\omega'}\tilde{\rho}_{\omega+\omega'}
   +\tilde{\rho}_{\omega}\hat{\rho}_{\omega'}\hat{\rho}_{\omega+\omega'}
   +\tilde{\rho}_{\omega}\hat{\rho}_{\omega'}\hat{\rho}_{\omega+\omega'},
\end{multline}
where,
\begin{equation}
  \tilde{\rho}_{\omega}=\sum_{n}\cos\omega \theta_n,\quad
  \hat{\rho}_{\omega}=\sum_{n}\sin\omega \theta_n.
\end{equation}

As we are considering even distributions of $\theta_n$, sums
involving the sines vanish, i.e. $\hat{\rho}_\omega=0$ for any
$\omega$. Hence, the sum is reduce to,
\begin{equation}
  \sum_{mnk}\cos\left(\omega \theta_{mn}+\omega' \theta_{kn}\right)=
  \tilde{\rho}_{\omega}\tilde{\rho}_{\omega'}\tilde{\rho}_{\omega+\omega'}.
\end{equation}

% ================================================
\section{Incompressible liquid of eigenvalues}\label{app:elst}

Consider the static equilibrium problem for a system of particles
with the pairwise interaction potential $\varphi (\lambda_{mn})$
given by \eqref{potential} where $\lambda_{mn}$ is the distance
between particles with numbers $m$ and $n$. The structure of the
potential is such that at small distance the interaction has
repulsive character while at large separation it is attractive. Thus
we may expect that $N$ particles interacting with such a potential
should have a static equilibrium position. Let us further assume
that this arrangement is compact for large $N$, i.e. the eigenvalue
density vanishes \emph{exactly} outside some finite region. Let us
assume this region to be connected of size $\Lambda$. The condition
of equilibrium requires that in the presence of non-zero density the
potential $\Phi (\lambda)$ defined as
\begin{equation}
  \Phi(\lambda)=\int_{-\Lambda/2}^{\Lambda/2}\dd
  \eta\,\rho(\eta)\varphi(\lambda-\eta),
\end{equation}
is constant $\Phi(\lambda)\equiv\Phi_0$. Due to this condition the
total energy is given by,
\begin{equation}
  E=\int\dd\lambda\,\rho(\lambda)\Phi(\lambda)=N\Phi_0,
\end{equation}
where we used the normalisability condition for the density.

Let us find the size of the distribution in the approximation of
incompressible condensate. For this let us consider the Taylor
expansion coefficients of the potential in the vicinity of origin:
\begin{equation}
  \Phi(\lambda)\equiv\Phi_0=\Phi(0)+\Phi'(0)\lambda+
  \ft12\Phi''(0)\lambda^2+\dots
\end{equation}

The zeroth term should give just $\Phi_0$,
\begin{equation}\label{phi0}
  \Phi_0=\rho_*\int_{-\Lambda/2}^{\Lambda/2}\dd\eta\,\varphi(\eta),
\end{equation}
where $\rho_*$ is an average value of the density, in present
approximation we can replace it by,
\begin{equation}\label{rho*}
  \rho_*=\frac{N}{\Lambda}.
\end{equation}

The first term in the expansion vanishes automatically due to the
symmetry of the distribution. Consider the third term:
\begin{equation}
  \Phi''_0=\rho_*\int_{-\Lambda/2}^{\Lambda/2}\dd\eta\,
  \rho(\eta)\varphi(\eta)=2\rho_* \varphi'(\Lambda/2).
\end{equation}
The vanishing of $\Phi''_0$, which is zero pressure condition at the
origin allows one to find the size of the condensate $\Lambda$,
\begin{equation}\label{Lambda-cond}
  \varphi'(\Lambda/2)=0.
\end{equation}

Now, plugging the the potential \eqref{potential} into into Eq.
\eqref{Lambda-cond} we get:
\begin{equation}
  \Lambda=2\sqrt{\mu_{+}\mu_{-}},\quad
  \rho_{*}=\frac{N}{2\sqrt{\mu_{+}\mu_{-}}},
\end{equation}
i.e. the eigenvalue density is described by,
\begin{equation}\label{density}
  \rho(\lambda)=
  \begin{cases}
    \frac{N}{2\sqrt{\mu_{+}\mu_{-}}},& -\sqrt{\mu_{+}\mu_{-}}<\lambda<
    \sqrt{\mu_{+}\mu_{-}}\\
    0, & |\lambda|\geq \sqrt{\mu_{+}\mu_{-}}.
  \end{cases}
\end{equation}

All this leads to the following result for the ``energy'',
\begin{equation}\label{final-e}
  4N^2\E=-4 N^2
  \left(\sqrt{\frac{\mu _+}{\mu _-}} \tan
   ^{-1}\left(\sqrt{\frac{\mu _-}{\mu _+}}\right)+\sqrt{\frac{\mu _-}{\mu _+}}\tan
   ^{-1}\left(\sqrt{\frac{\mu _+}{\mu _-}}\right)
  \right).
\end{equation}

% ======================================
\section{Entropy of random spin states}

Let us compute the number of the states of spin $s$ of a set of $L$
spin $1/2$ states. It is instructive to compare the entropy of this
set of random spin states with the one of the matrix model.

The problem is related to one of computation of the multiplicity
$\nu^{(L)}_s$ of the irreducible representation of spin $s$ in the
expansion of the product of $L$ spin-$1/2$ representations:
\begin{equation}\label{expan}
  \underbrace{\mathbf{1/2}\times \mathbf{1/2}\times\dots
  \mathbf{1/2}}_{L-\text{times}}=\sum_s
  \nu^{(L)}_s \mathbf{s},
\end{equation}
where we use the boldface letters $\mathbf{s}$ to denote the irreducible
representations of spin $s$.

To find the expansion \eqref{expan}, let us observe that the product
of an irreducible representation of $\mathbf{s}$ with $\mathbf{1/2}$
results in
\begin{equation}\label{arrows}
  \mathbf{1/2}\times\mathbf{s}=\mathbf{(s+1/2)}+\mathbf{(s-1/2)}.
\end{equation}

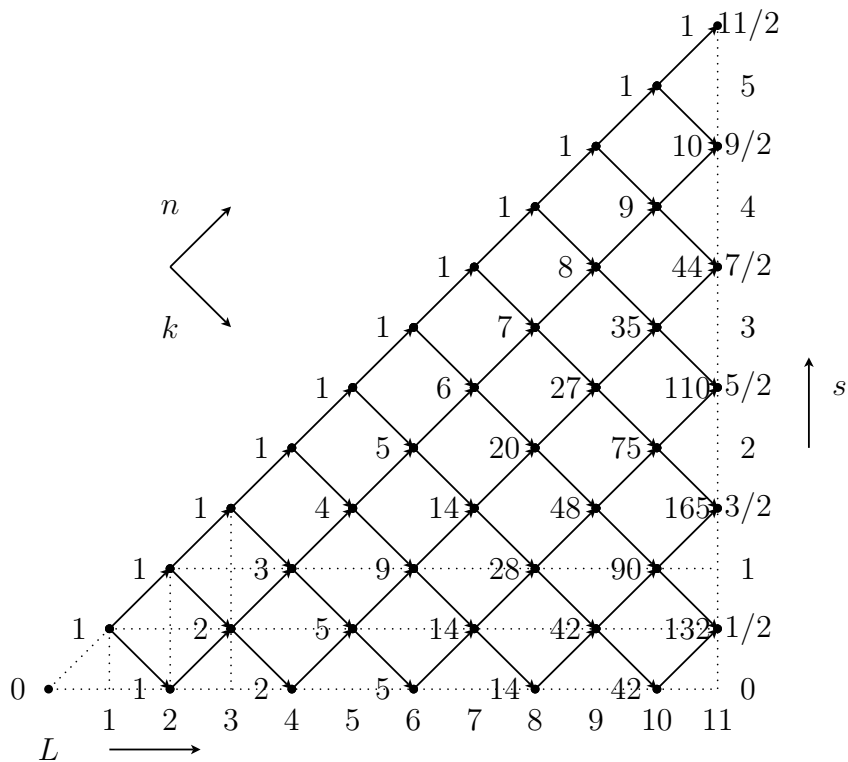
\begin{figure}[h]\label{fig:triangle}
\begin{center}
\psset{xunit=0.8mm,yunit=0.8mm,runit=0.8mm}
\psset{linewidth=0.3,dotsep=1,hatchwidth=0.3,hatchsep=1.5,shadowsize=1}
\psset{dotsize=0.7 2.5,dotscale=1 1,fillcolor=black}
\psset{arrowsize=1 2,arrowlength=1,arrowinset=0.25,tbarsize=0.7
5,bracketlength=0.15,rbracketlength=0.15}
\begin{pspicture}(0,0)(110,110)
\psline{->}(10,20)(20,10) \psdots[](10,20) (20,10)
\psline{->}(20,10)(30,20) \psdots[](20,10) (30,20)
\psline{->}(40,10)(50,20) \psdots[](40,10) (50,20)
\psline{->}(30,20)(40,10) \psdots[](30,20) (40,10)
\psline{->}(80,10)(90,20) \psdots[](80,10) (90,20)
\psline{->}(60,10)(70,20) \psdots[](60,10) (70,20)
\psline{->}(50,20)(60,10) \psdots[](50,20) (60,10)
\psline{->}(70,20)(80,10) \psdots[](70,20) (80,10)
\psline{->}(20,30)(30,40) \psdots[](20,30) (30,40)
\psline{->}(30,40)(40,50) \psdots[](30,40) (40,50)
\psline{->}(10,20)(20,30) \psdots[](10,20) (20,30)
\psline{->}(40,50)(50,60) \psdots[](40,50) (50,60)
\psline{->}(50,60)(60,70) \psdots[](50,60) (60,70)
\psline{->}(60,70)(70,80) \psdots[](60,70) (70,80)
\psline{->}(70,80)(80,90) \psdots[](70,80) (80,90)
\psline{->}(30,20)(40,30) \psdots[](30,20) (40,30)
\psline{->}(40,30)(50,40) \psdots[](40,30) (50,40)
\psline{->}(50,40)(60,50) \psdots[](50,40) (60,50)
\psline{->}(60,50)(70,60) \psdots[](60,50) (70,60)
\psline{->}(70,60)(80,70) \psdots[](70,60) (80,70)
\psline{->}(80,70)(90,80) \psdots[](80,70) (90,80)
\psline{->}(20,30)(30,20) \psdots[](20,30) (30,20)
\psline{->}(30,40)(40,30) \psdots[](30,40) (40,30)
\psline{->}(50,60)(60,50) \psdots[](50,60) (60,50)
\psline{->}(40,50)(50,40) \psdots[](40,50) (50,40)
\psline{->}(50,40)(60,30) \psdots[](50,40) (60,30)
\psline{->}(60,50)(70,40) \psdots[](60,50) (70,40)
\psline{->}(40,30)(50,20) \psdots[](40,30) (50,20)
\psline{->}(60,70)(70,60) \psdots[](60,70) (70,60)
\psline{->}(70,80)(80,70) \psdots[](70,80) (80,70)
\psline{->}(80,30)(90,20) \psdots[](80,30) (90,20)
\psline{->}(90,80)(100,70) \psdots[](90,80) (100,70)
\psline{->}(80,70)(90,60) \psdots[](80,70) (90,60)
\psline{->}(70,60)(80,50) \psdots[](70,60) (80,50)
\psline{->}(60,30)(70,20) \psdots[](60,30) (70,20)
\psline{->}(100,70)(110,60) \psdots[](100,70) (110,60)
\psline{->}(90,40)(100,30) \psdots[](90,40) (100,30)
\psline{->}(90,60)(100,50) \psdots[](90,60) (100,50)
\psline{->}(70,20)(80,30) \psdots[](70,20) (80,30)
\psline{->}(80,50)(90,40) \psdots[](80,50) (90,40)
\psline{->}(70,40)(80,30) \psdots[](70,40) (80,30)
\psline{->}(50,20)(60,30) \psdots[](50,20) (60,30)
\psline{->}(100,30)(110,40) \psdots[](100,30) (110,40)
\psline{->}(100,50)(110,40) \psdots[](100,50) (110,40)
\psline{->}(100,30)(110,20) \psdots[](100,30) (110,20)
\psline{->}(90,20)(100,10) \psdots[](90,20) (100,10)
\psline{->}(100,10)(110,20) \psdots[](100,10) (110,20)
\psline{->}(90,60)(100,70) \psdots[](90,60) (100,70)
\psline{->}(90,40)(100,50) \psdots[](90,40) (100,50)
\psline{->}(80,30)(90,40) \psdots[](80,30) (90,40)
\psline{->}(90,20)(100,30) \psdots[](90,20) (100,30)
\psline{->}(60,30)(70,40) \psdots[](60,30) (70,40)
\psline{->}(80,50)(90,60) \psdots[](80,50) (90,60)
\psline{->}(70,40)(80,50) \psdots[](70,40) (80,50)
\psline{->}(100,50)(110,60) \psdots[](100,50) (110,60)
\psline{->}(80,90)(90,100) \psdots[](80,90) (90,100)
\psline{->}(80,90)(90,80) \psdots[](80,90) (90,80)
\psline{->}(100,90)(110,80) \psdots[](100,90) (110,80)
\psline{->}(100,70)(110,80) \psdots[](100,70) (110,80)
\psline{->}(90,80)(100,90) \psdots[](90,80) (100,90)
\psline{->}(90,100)(100,90) \psdots[](90,100) (100,90)
\psline{->}(90,100)(100,110) \psdots[](90,100) (100,110)
\psline{->}(100,90)(110,100) \psdots[](100,90) (110,100)
\psline{->}(100,110)(110,100) \psdots[](100,110) (110,100)
\psline{->}(100,110)(110,120) \psdots[](100,110) (110,120)
\psline[linestyle=dotted](0,10)(10,20)
\psdots[linestyle=dotted](0,10) (10,20)
\psline[linestyle=dotted](10,20)(10,10)
\psbezier[linestyle=dotted](20,30)(20,30)(20,30)(20,30)
\psline[linestyle=dotted](20,30)(20,10)
\psline[linestyle=dotted](30,40)(30,10)
\psline[linestyle=dotted](110,10)(110,120)
\psline[linestyle=dotted](0,10)(110,10)
\psline[linestyle=dotted](110,20)(10,20)
\psbezier[linestyle=dotted](0,30)(0,30)(0,30)(0,30)
\psline[linestyle=dotted](20,30)(110,30) \rput(130,60){$s$}
\psline{->}(125,50)(125,65) \psline{->}(10,0)(25,0) \rput(0,0){$L$}
\rput(10,5){1} \rput(20,5){2} \rput(30,5){3} \rput(40,5){4}
\rput(50,5){5} \rput(60,5){6} \rput(70,5){7} \rput(80,5){8}
\rput(90,5){9} \rput(100,5){10} \rput(110,5){11} \rput(115,20){1/2}
\rput(115,10){0} \rput(115,30){1} \rput(115,40){3/2}
\rput(115,50){2} \rput(115,60){5/2} \rput(115,70){3}
\rput(115,80){7/2} \rput(115,90){4} \rput(115,100){9/2}
\rput(115,110){5} \rput(115,120){11/2} \rput(5,20){1}
\rput(15,30){1} \rput(25,40){1} \rput(35,50){1} \rput(45,60){1}
\rput(55,70){1} \rput(65,80){1} \rput(75,90){1} \rput(85,100){1}
\rput(95,110){1} \rput(105,120){1} \rput(15,10){1} \rput(25,20){2}
\rput(35,10){2} \rput(35,30){3} \rput(45,40){4} \rput(55,50){5}
\rput(65,60){6} \rput(75,70){7} \rput(85,80){8} \rput(95,90){9}
\rput(105,100){10} \rput(45,20){5} \rput(55,30){9} \rput(65,40){14}
\rput(75,50){20} \rput(85,60){27} \rput(95,70){35} \rput(105,80){44}
\rput(55,10){5} \rput(65,20){14} \rput(75,30){28} \rput(85,40){48}
\rput(95,50){75} \rput(105,60){110} \rput(75,10){14}
\rput(85,20){42} \rput(95,30){90} \rput(105,40){165}
\rput(95,10){42} \rput(105,20){132} \psline{->}(20,80)(30,90)
\psline{->}(20,80)(30,70) \rput(20,90){$n$} \rput(20,70){$k$}
\rput(-5,10){0}
\end{pspicture}
% ==========================================================
\end{center}
\caption{Triangle of $L$ products of  $\mathbf{1/2}$
  representations. The horizontal levels count (in the direction $s$)
  the spin of irreducible
  representations. The vertical dotted lined correspond to the length
  of the product. Thus the maximal length in this picture corresponds
  to the product of 11 $\mathbf{1/2}$. The number to the left of each
  dot represents the number of distinct ways the dot can be reached
  from the ``zero point'' following the arrows. This number gives the
  degeneracy of the representation given by the level in the
  decomposition of the product of $\mathbf{1/2}$ in the number of
  factors is given by the horizontal coordinate $L$.}
\end{figure}

Consequent multiplication by $\mathbf{1/2}$ in the l.h.s. of
spin $\mathbf{1/2}$ representation in\eqref{expan} can be represented
by the diagram \ref{fig:triangle}, where each dot represent an
irreducible representation of su(2). According to \eqref{arrows}
multiplication by an additional factor of $\mathbf{1/2}$ gives rise to
a representation of representations of spin differing by $\pm1/2$ with
factor one each. On the picture this is represented  by arrows. Thus
the the total number of factors of spin $s$ one gets after
multiplication of $L$ $\mathbf{1/2}$ factors is given by the number of
distinct paths of length $L$ one can reach the level corresponding to
the spin $s$ from the left-most position following the arrows.

Thus, the problem is reduced to the computation of the number of
distinct paths to reach the level $s$. To do this let us note that
the representations form families corresponding to straight lines
starting from the bottom. Let us assign a number $k=0,\dots, L$ to
each line. The multiplicity of irreducible representations found on
a particular line satisfy simple properties. For example the
multiplicities on the zeroes line are all equal to one. On the line
number one the multiplicities are given by the sum of multiplicities
from the zeroes line up to the next number of the sequence term. In
general, the multiplicity on the $k$th line is connected to ones on
the $(k-1)$th one by the following recurrence relation
\begin{equation}\label{recurr}
  \nu_k^n=\sum_{l=2}^{n+1}\nu_{k-1}^l,\qquad \nu_k^1=\nu_{k-1}^2.
\end{equation}

The recurrence relation \eqref{recurr} can be solved which leads to
the following expression fo the multiplicities,
\begin{equation}
  \nu^n_k=n\frac{(n+2k-1)!}{k!(n+k)!}.
\end{equation}

Now taking into account that $k$ and $n$ can be expressed in terms of
the number of spins $L$ and and the spin $s$ as follows\footnote{The
 geometrical meaning of $n$ is the dimension of corresponding spin
 representation.}
\begin{equation}
  L=n+2k-1,\qquad s=L/2-k=(n-1)/2,
\end{equation}
one can rewrite the multiplicities in the following form
\begin{equation}\label{nr-of-spins}
  \nu^{(L)}_s=(2s+1)\frac{L!}{(L/2-s)!(L/2+s+1)!}.
\end{equation}

The check that the sum of the dimensions all irreducible
representations taking into account the multiplicities is indeed
\begin{equation}
  \sum_s (2s+1)\nu^{(L)}_s=2^L,
\end{equation}
is left to the reader.

In the limit of large $L$ one can use the Stirling's approximation
to the factorials. This gives the log number of $L$ random spin
states of spin $s$ to be,
\begin{equation}\label{random-spin-entr}
  S_{\rm random~spin}=L\ln L-\ft12\left\{
  (L-2s)\ln (L-2s)+(L+2s)\ln(L+2s)\right\}.
\end{equation}
It is useful to compare this quantity to the spin entropy of the
matrix oscillator in the main body of the paper.
% ------------------------------------------------------------

%% -----------------------------------------------------------
%% -----------------------------------------------------------
%\bibliographystyle{hunsrt}
%\bibliography{ad1}

\end{document}